\documentclass[aps,pre,twocolumn,footinbib,tightenlines,superscriptaddress,final,letterpaper,longbibliography,nofootinbib]{revtex4-2}

\usepackage[T1]{fontenc}
\usepackage[utf8]{inputenc}

\usepackage{amsmath, gensymb, dsfont}
\usepackage{physics}
\usepackage{hyperref}
\usepackage{graphicx, float}
\usepackage[caption=false]{subfig}
\usepackage{xcolor}

\definecolor{monbbleu}{RGB}{76, 114, 176}

\hypersetup{unicode=false, pdftoolbar=true, pdfmenubar=true, pdffitwindow=false, pdfstartview={FitH}, pdfnewwindow=true, pdftitle={}, pdfauthor={}, pdfsubject={}, pdfkeywords={}, colorlinks=true, allcolors=monbbleu}

\newcommand{\posterior}{\textit{a posteriori}}
\newcommand{\prior}{\textit{a priori}}
\newcommand{\comment}[1]{}

\DeclareMathOperator{\TruncGamma}{TruncGamma}

\DeclareMathOperator*{\argmax}{argmax}

\begin{document}

\title{Hypergraph reconstruction from noisy pairwise observations}

\author{Simon Lizotte}
\affiliation{D\'epartement de physique, de g\'enie physique et d'optique, Universit\'e Laval, Qu\'ebec (Qu\'ebec), Canada G1V 0A6}%
\affiliation{Centre interdisciplinaire en mod\'elisation math\'ematique, Universit\'e Laval, Qu\'ebec (Qu\'ebec), Canada G1V 0A6}%

\author{Jean-Gabriel Young}
\affiliation{D\'epartement de physique, de g\'enie physique et d'optique, Universit\'e Laval, Qu\'ebec (Qu\'ebec), Canada G1V 0A6}%
\affiliation{Department of Mathematics and Statistics, University of Vermont, Burlington, VT 05405, USA}
\affiliation{Vermont Complex Systems Center, University of Vermont, Burlington, VT 05405, USA}

\author{Antoine Allard}
\affiliation{D\'epartement de physique, de g\'enie physique et d'optique, Universit\'e Laval, Qu\'ebec (Qu\'ebec), Canada G1V 0A6}%
\affiliation{Centre interdisciplinaire en mod\'elisation math\'ematique, Universit\'e Laval, Qu\'ebec (Qu\'ebec), Canada G1V 0A6}%
\affiliation{Vermont Complex Systems Center, University of Vermont, Burlington, VT 05405, USA}

\begin{abstract}
    The network reconstruction task aims to estimate a complex system's structure from various data sources such as time series, snapshots, or interaction counts.
    Recent work has examined this problem in networks whose relationships involve precisely two entities---the pairwise case.
    Here we investigate the general problem of reconstructing a network in which higher-order interactions are also present.
    We study a minimal example of this problem, focusing on the case of hypergraphs with interactions between pairs and triplets of vertices, measured imperfectly and indirectly.
    We derive a Metropolis-Hastings-within-Gibbs algorithm for this model and use the algorithms to highlight the unique challenges that come with estimating higher-order models.
    We show that this approach tends to reconstruct  empirical and synthetic networks more accurately than an equivalent graph model without higher-order interactions.
\end{abstract}

\maketitle

\section{Introduction}

Networks are a convenient model for the intricate structure of complex systems, in which interactions between any pair of the system's constituting elements can be directly interpreted as edges between the corresponding vertices of a graph.
In typical network analyses, these pairwise interactions will initially be unknown as we cannot observe them directly; one must instead define a model of what is and is not an interaction and put this model to the data to identify the relevant network.
For instance, we might define a pollinator and a plant species as interacting if a pollinator prefers a particular species over others.
This definition will then let us infer a plant-pollinator interaction network by observing how often each pollinator visits each plant and processing the data with an appropriate statistical model~\cite{basilio_2006_YearlongPlantpollinatorNetwork,young_2021_ReconstructionPlantPollinator}.

Numerous methods have been proposed to perform this critical step of the network analysis process, commonly called graph reconstruction.
They span a broad range of statistical and machine learning techniques and are often tailored to the specific field for which they have been developed~\cite{brugere_2018_NetworkStructureInference}.
Gene regulatory networks, for instance, have been  reconstructed with methods ranging from random forests~\cite{huynh-thu_2010_InferringRegulatoryNetworks} to methods based on Pearson correlation in temporal windows~\cite{specht_2017_LEAPConstructingGene} or in ordinary differential equations~\cite{matsumoto_2017_SCODEEfficientRegulatory}.
Bayesian frameworks based on genomic features~\cite{jansen_2003_BayesianNetworksApproach} or random-walk-based algorithms~\cite{lei_2013_NovelLinkPrediction} have been used to estimate protein-protein interaction networks; while brain networks have been measured with a broad range of methods like cross-frequency phase synchronization~\cite{cai_2018_ReconstructionFunctionalBrain}, Granger causality~\cite{hlavackova-schindler_2007_CausalityDetectionBased}, and matrix-regularized network learning frameworks~\cite{qiao_2016_EstimatingFunctionalBrain}.
More general methods have also been developed to reconstruct a broad range of datasets~\cite{peixoto_2018_ReconstructingNetworksUnknown, runge_2018_CausalNetworkReconstruction, newman_2018_NetworkStructureRich, kramer_2009_NetworkInferenceConfidence, young_2021_BayesianInferenceNetwork}.

While convenient, graphs are fundamentally limited to encoding dyadic connections because higher-order interactions aren't always reducible to a set of pairwise ties~\cite{battiston_2021_PhysicsHigherorderInteractions, battiston_2021_PhysicsHigherorderInteractions, bick_2022_WhatAreHigherorder}. For example, empirical evidence shows that accounting for such higher-order interactions  can enhance models of cortical dynamics~\cite{yu_2011_HigherOrderInteractionsCharacterized}, of the diversity of species in natural communities~\cite{mayfield_2017_HigherorderInteractionsCapture, bairey_2016_HighorderSpeciesInteractions, grilli_2017_HigherorderInteractionsStabilize}, and of social group formation~\cite{milojevic_2014_PrinciplesScientificResearch}.
If they are to reap the benefits of such representations, network science methods should always be able to handle higher-order interactions whenever dyadic relationships are insufficient.

There has been significant recent progress in adapting the network science methods to higher-order representations~\cite{battiston_2020_NetworksPairwiseInteractions}, but the  higher-order reconstruction problem has no fully satisfactory solution to date.
For instance, direct generalizations of dyadic algorithms are impractical due to the extraordinary amount of data they require to function~\cite{battiston_2020_NetworksPairwiseInteractions}---a network of $n$ vertices can support up to $2^n$ hyperedges so naively measuring every edge becomes rapidly infeasible with growing $n$.
Recent work has addressed this issue partially by using pairwise data to make inferences about possible higher-order structures~\cite{young_2021_HypergraphReconstructionNetwork} or  filters on incomplete hyperedge data~\cite{musciotto_2021_DetectingInformativeHigherorder}.
However, no method to date can simultaneously handle reconstruction and noise in the pairwise measurements.

This paper introduces a Bayesian framework to infer higher-order structural interactions from imperfect pairwise measurements. We study a minimal example of this problem, focusing on the case of hypergraphs with interactions between pairs and triplets of vertices, measured imperfectly and indirectly. Instead of providing a point estimate, this framework offers a distribution of the possible hypergraphs compatible with all the available observations.  The range of structures provided by this distribution allows us to compute error bars for various network measurements and the outcomes of network processes. We also present a network modeling approach that encodes the projection of hyperedges as different types of pairwise interactions to analyze the importance of the correlation induced by these higher-order interactions. Finally, we compare the reconstruction accuracy of these two frameworks on synthetic observations generated from various synthetic and empirical hypergraphs.

\section{Methods}

Let us assume that we possess some measurements $X=[x_{ij}]_{i,j=1,\ldots,n}$ of the pairwise interactions of the units of a complex system composed of $n$ elements. In general reconstruction problems, these observations could take on many forms, such as time series correlation of brain regions~\cite{stam_2004_FunctionalConnectivityPatterns} or the direct observation of the presence (or absence) of edges in a networked system~\cite{peixoto_2018_ReconstructingNetworksUnknown}, to name only two examples.
To keep our presentation of the methods concrete, we will focus on the case where $x_{ij}$ is an integer number of observed interactions for vertices $i$ and $j$.
Our objective is to infer the interactions in a hidden latent structure $\mathcal{S}$ under the assumption that these interactions shape the observed behavior of the system (i.e., the measurements). This latent structure could be any type of structural representation such as graphs, simplicial complexes, or hypergraphs.

We expect the observation data to be noisy, meaning that remeasuring the system will lead to different values $X$. We also expect that similar (different) interactions in $\mathcal{S}$ could lead to very different (similar) measurements. For instance, two pairwise observations $x_{ij}$ and $x_{rs}$ could be identical even if the pair $(i,j)$ interacts in $\mathcal{S}$ while $(r,s)$ does not.
To account for these fluctuations, we develop a Bayesian inference framework, a fully probabilistic approach producing a probability distribution over the different structures $\mathcal{S}$ compatible with the data $X$.

\subsection{Data model}

We first specify the likelihood $P(X|\mathcal{S}, \mu)$, which expresses how the observations $X$ are related to the latent structure $\mathcal{S}$ and any additional parameters of the observation processes $\mu$. We assume that the structure $\mathcal{S}$ encodes three types of symmetrical interactions: each pair $(i,j)$ can interact weakly $\ell_{ij}=1$, interact strongly $\ell_{ij}=1$ or not interact $\ell_{ij}=0$. For instance, measurements $X$ of a social network could be the number of conversations recorded between acquaintances ($\ell_{ij}=1$), friends ($\ell_{ij}=2$) or strangers ($\ell_{ij}=0$).

\begin{figure}
    \centering
    \includegraphics[width=\linewidth]{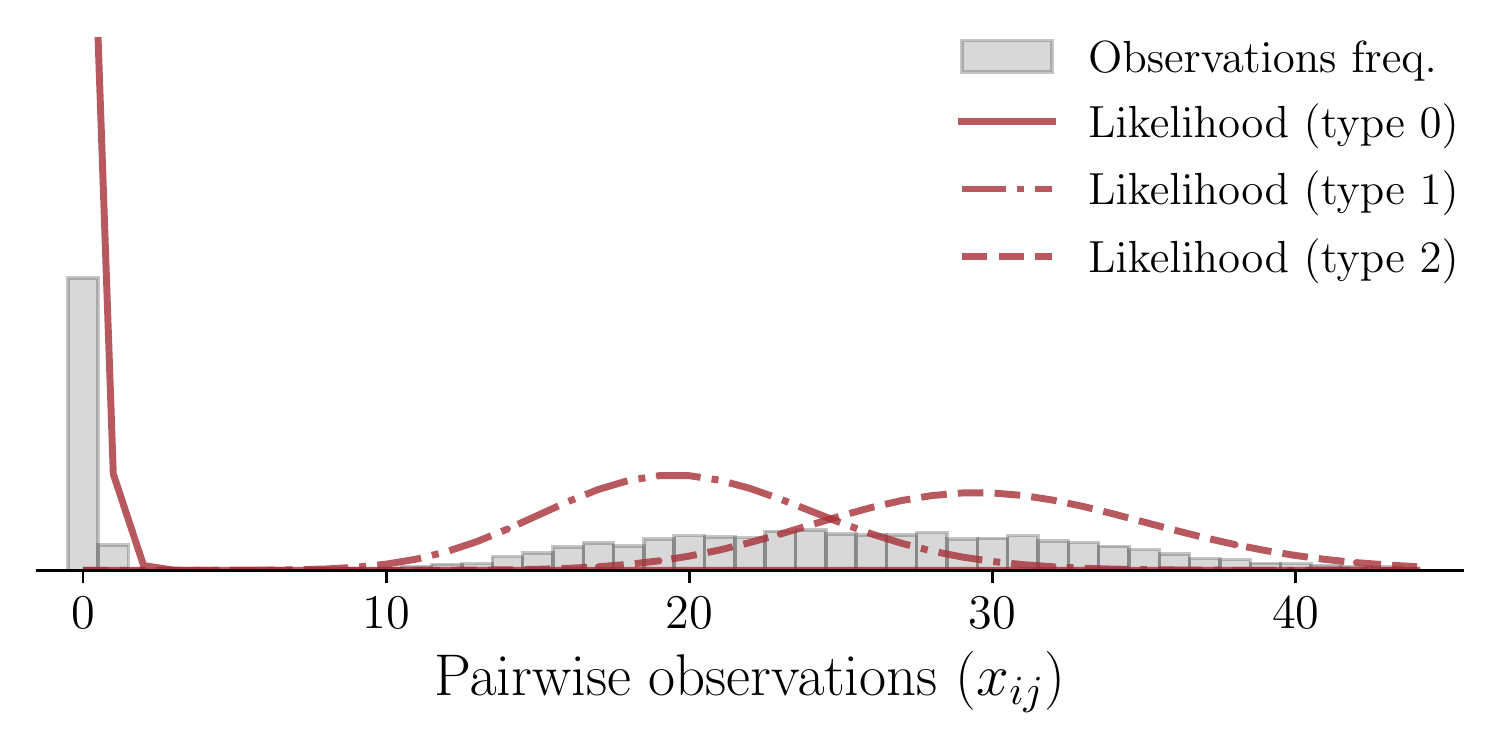}
    \caption{Illustration of a typical distribution of pairwise interactions $X$ produced by the data model. The frequencies of the pairwise interactions are shown in gray.  The contribution of each type of interaction to the likelihood is shown in red.}
    \label{fig:mixture_dist}
\end{figure}

For a fairly broad range of measurement processes, it will often be reasonable to model the observed number of interactions $x_{ij}$  between vertices $i$ and $j$ with conditionally independent Poisson point processes. In such processes, the observation $x_{ij}$ is only determined by its associated type of interaction $\ell_{ij}$ and average $\mu_{\ell_{ij}}$, leading  to the likelihood
\begin{align}
    P(X|\mathcal{S}, \mu) = \prod_{i<j} \frac{\mu_{\ell_{ij}}^{x_{ij}}}{x_{ij}!}e^{-\mu_{\ell_{ij}}},
    \label{eq:likelihood}
\end{align}
where $\mu = (\mu_0, \mu_1, \mu_2)$. Figure~\ref{fig:mixture_dist} illustrates the distribution of pairwise observations modeled by Eq.~\eqref{eq:likelihood}.
This model will only be appropriate if the errors on two distinct measurements $x_{ij}$ and $x_{rs}$ are not correlated, and every $x_{ij}$ is the outcome of numerous independent observations of an ongoing measurement process with constant success rate.
We make these assumptions to provide a simple illustration of our inference framework, but we stress that it is general enough to account for more general and diverse types of data, distributions, and structure.

\subsection{Structural models}
\label{ssec:s_models}

The next step is to specify the latent structural model $P(\mathcal{S}|\phi)$, which is a prior probability on each interaction  $\ell_{ij}$ conditioned on some additional parameters collectively denoted by $\phi$. This distribution encodes our hypothesis on the structure of interactions of the system before we make any measurements. For instance, we might expect person $i$ to be more likely to develop a friendship with person $j$ than with person $k$ because $i$ and $j$ live in the same neighborhood.

To highlight the role of latent higher-order interactions in the reconstruction procedure (or lack thereof), we consider two models for the structure $\mathcal{S}$: a hypergraph model ($\mathcal{S}=H$) and a categorical-edge model with a simpler graph structure ($\mathcal{S}=G$).

\subsubsection{Hypergraph model}

We define the hypergraph structure $H=(V, E, T)$ as a set of vertices $V$ with 2-edges $E$ and 3-edges $T$. We limit the size of the hyperedges to 3 for the sake of simplicity, although larger hyperedges could easily be considered by adapting the data model in Eq.~\eqref{eq:likelihood} accordingly.
We opt for a simple hypergraph model in which the existence of each hyperedge is conditionally independent from the others.
Denoting as $p$ ($q$) the probability of existence of 3-edges (2-edges), the probability of $H$ is
\begin{align}
    P(H|\phi_H) = q^{h_1} (1-q)^{\binom{n}{2}-h_1} p^{h_2} (1-p)^{\binom{n}{3}-h_2},
    \label{eq:phg_prior}
\end{align}
where $\phi_H=\{p,q\}$ are the parameters, $h_1=|E|$ is the number of 2-edges and $h_2=|T|$ is the number of 3-edges.

We connect this structure to the data model by assigning a type $\ell_{ij}$ to each pair of vertices as
\begin{align}
    \ell_{ij}= \begin{cases}
        2 & \text{if }(i,j) \in \Delta,\\
        1 & \text{if }(i,j) \in E \text{ and if } (i,j)\not\in \Delta,\\
        0 & \text{otherwise.}
    \end{cases}
    \label{eq:phg_labels}
\end{align}
where $\Delta$ is the set of pairs covered by a 3-edge
\begin{align}
    \Delta = \{ (i, j)\ |\ \exists\ k \text{ s.t. } (i,j,k) \in T \}.
\end{align}

To make further progress, we must make a few arbitrary choices since the full model---the joint distribution of the data and latent structure---can be re-parametrized in ways that do not affect the distribution over labels and, therefore, over data.
These symmetries will cause identifiability problems when we use the model to make inferences about latent hypergraphs, so we address them immediately.

\begin{figure}
    \centering
    \includegraphics[width=5cm]{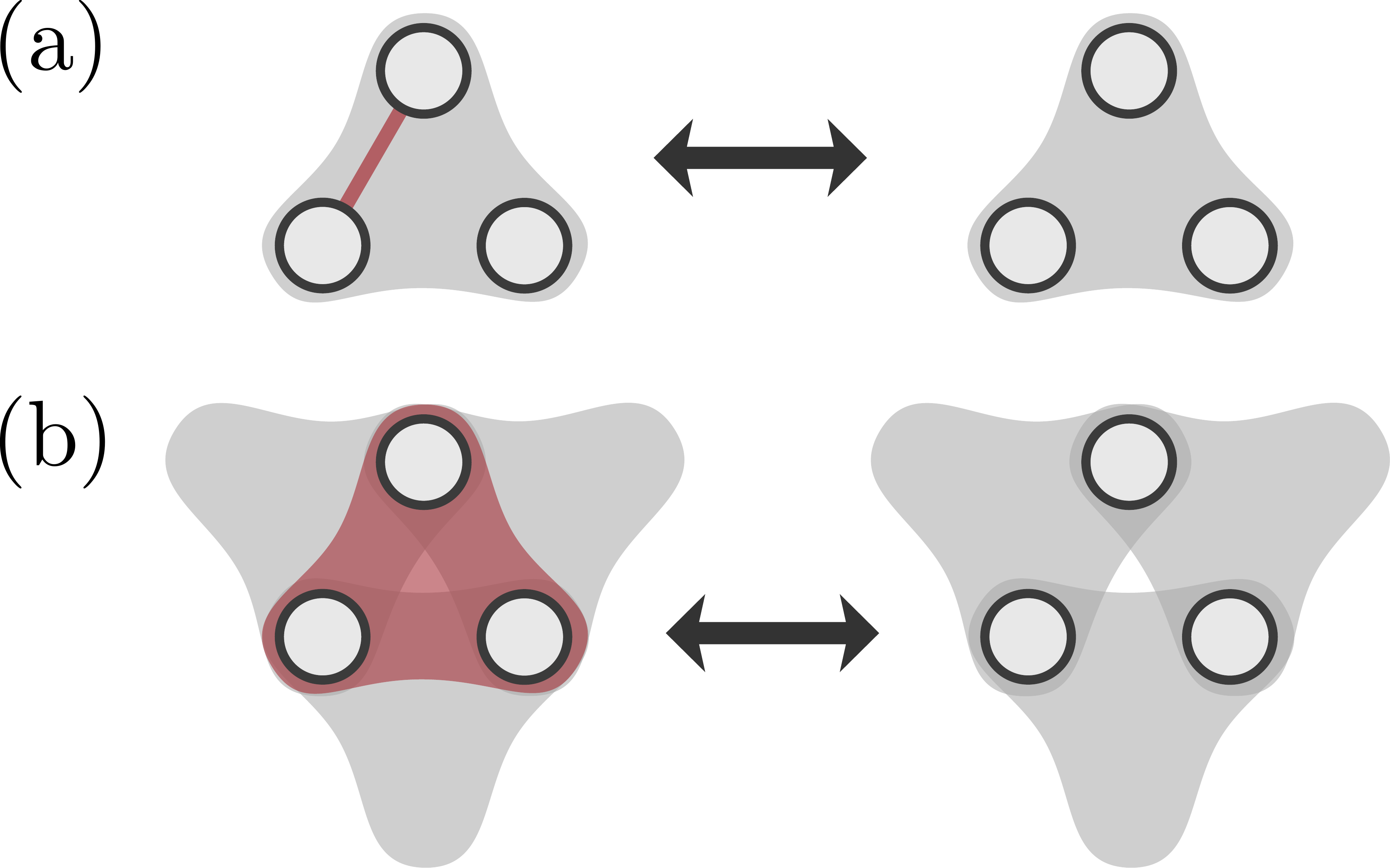}
    \caption{Examples of structural configurations with hidden edges. The presence or absence of the (a) 2-edge and (b) 3-edge shown in red does not alter the type of interaction $\ell_{ij}$ of the vertices, which is the same for all configurations. Hence, the likelihood in  Eq.~\eqref{eq:likelihood} has the same value, and we say that these red hyperedges are \textit{hidden} by the other 3-edges.}
    \label{fig:non_identifiability}
\end{figure}

First, since the mapping from hypergraph to labels is lossy, the presence of some edges can be \emph{hidden} by others.
For example, if vertices $i$ and $j$ are connected by both a 2-edge and a 3-edge, then the interaction will be considered of type $\ell_{ij} = 2$, as if the 2-edge did not exist---removing them does not affect the interaction type and consequently does not change the value of the likelihood given at Eq.~\eqref{eq:likelihood}. 3-edges can also hide other 3-edges, as depicted in Fig.~\ref{fig:non_identifiability}.
Hence, we must bear in mind that we will only be able to make inferences about ``visible'' edges.

Second, the full model is susceptible to label-switching and thus needs additional adjustments.
Indeed, while a non-interacting pair ($\ell_{ij}=0$) and a pair of vertices connected by a 2-edge ($\ell_{ij}=1$) are associated with different distributions of observations because they have distinct means $\mu_0$ and $\mu_1$, it is possible to change the structure $H$ and the parameters $\mu$ in a way that will not affect the overall likelihood of a dataset $X$.
This can be done by replacing every  non-interacting pair of $H$ by a 2-edge and vice-versa while also swapping the value of $\mu_0$ and $\mu_1$. We address this label-switching symmetry it in the standard way by imposing that $\mu_0<\mu_1$ or, equivalently, by thinking of non-interacting pairs as associated with a smaller expected number of interactions than interacting pairs.

The label $\ell_{ij}=2$ can also technically be exchanged with the labels $\ell_{ij}=0$ and $\ell_{ij}=1$, but because they are inherited from a latent hypergraph that correlates multiple pairs of vertices, the problem will only manifest itself in very specific situations.
Namely, every 2-edge has to belong to at least one triangle formed by two other 2-edges or projected 3-edges (this worst-case hypergraph is described in section~\ref{ssec:synthetic_inference}). Since a vanishing fraction of hypergraphs exhibit this specific configuration, imposing $\mu_1<\mu_2$ is unnecessary to disambiguate most configurations. That said, in practice, we found it useful to impose $\mu_0<\mu_2$. Type 1 and type 2 interactions are typically sparse, which means that type 0 interactions are dense. Non-interacting pairs could therefore seem to form many triangles and could be interpreted as the projection of 3-edges. Imposing $\mu_0<\mu_2$ avoids any confusion.

\subsubsection{Categorical-edge model}

Our second model involves
graph with categorical edges $G=(V, E_1, E_2)$ defined as a set of vertices $V$, of \emph{weak} edges $E_1$, and of \emph{strong} edges $E_2$. The types of interaction are then
\begin{align}
    \ell_{ij} = \begin{cases}
        2 & \text{if } (i,j) \in E_2,\\
        1 & \text{if } (i,j) \in E_1,\\
        0 & \text{otherwise.}
    \end{cases}
\end{align}
Much like in the hypergraph case, we adopt an agnostic model and assume \prior{} that the categorical edges are placed randomly according to a simple two-step generative process: strong edges are created independently with probability $q_2$ and weak edges are created independently in the remaining unconnected pairs with probability $q_1$
\begin{multline}
    P(G| \phi_G) = q_1^{m_1} (1-q_1)^{\binom{n}{2}-m_1-m_2} \\
    \times q_2^{m_2} (1-q_2)^{\binom{n}{2} - m_2},
    \label{eq:pes_prior}
\end{multline}
where $\phi_G=\{q_1,q_2\}$, $m_1=|E_1|$ and $m_2=|E_2|$ are the number of weak edges and strong edges respectively.

There are no hidden edges in this model but the label switching problem is now three-fold: $\ell_{ij}=0$ can be swapped with $\ell_{ij}=1$, but also $\ell_{ij}=0$ with $\ell_{ij}=2$ and $\ell_{ij}=1$ with $\ell_{ij}=2$. As for the hypergraph model, we address this issue by imposing $\mu_0<\mu_1<\mu_2$. Hence, we suppose that non-interaction pairs are less frequently measured than interactions and that weak interactions are less frequently measured than the strong ones.

\begin{figure*}
    \centering
    \includegraphics[width=.85\linewidth]{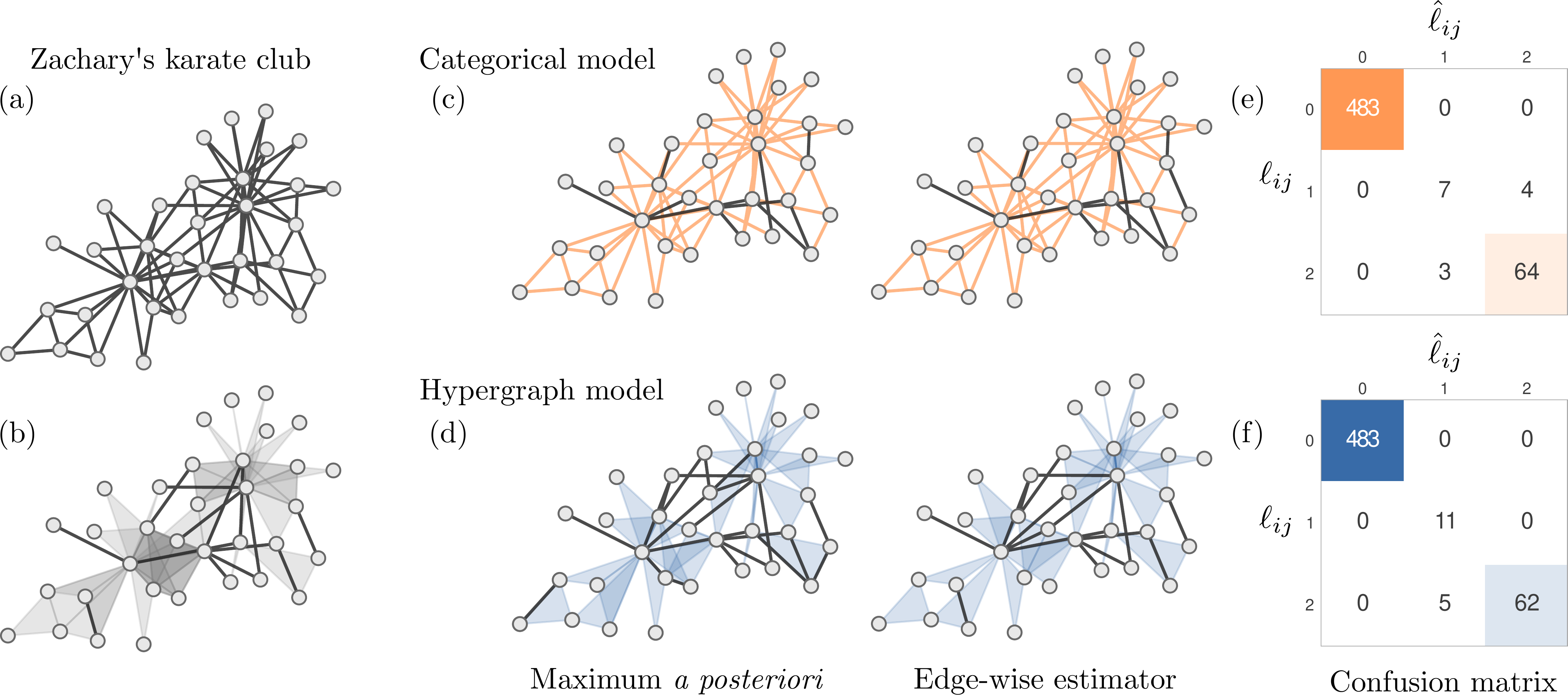}
    \caption{Example of the method inference process on a small dataset.
    (a) Original network of Zachary's karate club~\cite{zachary_1977_InformationFlowModel}.
    (b) Hypergraph representation of the Zachary's karate club, see main text.
    (c) Illustration of the structure corresponding to the estimators $\hat{\mathcal{S}}_\text{MAP}$ and $\hat{\mathcal{S}}_\text{EW}$ for the categorical-edge model. Strong edges are shown in orange.
    (d) Same as (c) but using the hypergraph model.
    (e) Confusion matrix built using the $\hat{S}_\text{MM}$ estimators for the interaction types.
    (f) Same as (e) but using the hypergraph model.
    The inference was done on synthetic observations generated using $\mu_0=0.01$, $\mu_1=20$ and $\mu_2=30$.  The Maximum \textit{a posteriori} (MAP) structure maximizes the posterior distributions [Eqs.~\eqref{eq:phg_posterior} and~\eqref{eq:pes_posterior}], while the average structure contains the edges and hyperedges that exist in at least half of the samples of the posterior distributions. The estimator for the type of interaction, noted $\hat{\ell}_{ij}$, is used to build the confusion matrix. It corresponds to the most likely type of interaction for vertices $i$ and $j$.}
    \label{fig:zachary_graphs}
\end{figure*}

\subsection{Posterior distributions}

Combining the quantities defined above, the Bayes formula yields
the posterior distribution $P(\mathcal{S}, \mu, \phi|X)$ of each structural model
\begin{align}
    P(\mathcal{S}, \mu, \phi|X) = \frac{P(X|\mathcal{S},\mu) P(\mathcal{S}|\phi)P(\mu, \phi)}{P(X)},
    \label{eq:bayes_model}
\end{align}
where $P(\mu, \phi)$ is a conjugate prior distribution (see Appendix~\ref{annex:priors} for details) and $P(X)$ is a normalization factor that need not be specified.

Combining Eqs.~\eqref{eq:phg_prior} and~\eqref{eq:pes_prior} with~\eqref{eq:bayes_model} yields the following posterior distributions
\begin{multline}
    P(H, \mu, \phi_H|X) = \frac{P(\mu, \phi)}{P(X)} q^{h_1} (1-q)^{\binom{n}{2}-h_1} \\
    \times p^{h_2} (1-p)^{\binom{n}{3}-h_2}
    \prod_{i<j} \frac{ (\mu_{\ell_{ij}})^{x_{ij}} }{x_{ij}!} e^{-\mu_{\ell_{ij}}}
    \label{eq:phg_posterior}
\end{multline}
and
\begin{multline}
    P(G, \mu, \phi_G|X) =  \frac{P(\mu, \phi)}{P(X)} q_1^{m_1} (1-q_1)^{\binom{n}{2}-m_1-m_2} \\
    \times q_2^{m_2} (1-q_2)^{\binom{n}{2} - m_2}
                    \prod_{i<j} \frac{ (\mu_{\ell_{ij}})^{x_{ij}} }{x_{ij}!} e^{-\mu_{\ell_{ij}}},
    \label{eq:pes_posterior}
\end{multline}
which both weight every structure-parameters tuple $(S,\mu,\phi)$ according to their compatibility with the observations $X$ and the prior probabilities.

Equations~\eqref{eq:phg_posterior}--\eqref{eq:pes_posterior} are not closed forms of known distributions,  with the main complication being due to the presence of edge labels $\ell_{ij}$ in the likelihood. Hence, any meaningful use of these posterior distributions will require the generation of samples from it, which in turn will be used to estimate statistics such as percentiles, the average and the variance of various functions $f(\mathcal{S}, \mu, \phi)$. To this end, we have derived a Metropolis-within-Gibbs algorithm whose details are discussed in Appendix.~\ref{annex:gibbs}. A C++/Python implementation is available at \url{https://github.com/DynamicaLab/hypergraph-bayesian-reconstruction}. The algorithm returns a series of tuples  $\{(S_t,\mu_t,\phi_t)\}_{t=1,...N}$ sampled in proportion to Eq.~\eqref{eq:bayes_model}, for either structural models.

\section{Results}

{\setlength{\tabcolsep}{1em}
\begin{table*}
    \centering
    \caption{Properties of the synthetic and empirical hypergraph datasets. The table shows the number of vertices ($n$), the number of type-1 and type-2 interactions, the relative reconstruction error ($\epsilon$) for both the categorical edges graph model and the hypergraph model, as well as the fraction of 2-edges that are hidden under a 3-edge ($E_\Delta$).  The relative reconstruction error ($\epsilon$) shown here is the median of the relative reconstruction error for 10 observation matrices generated with $(\mu_0=0.01, \mu_1=40, \mu_2=50)$.}
    \label{tab:real_data_results}
    \begin{tabular}{@{\extracolsep{4pt}}lcccccc@{}}
    \hline
    \hline
    \rule{0pt}{2ex}%
         Hypergraph & $n$ & \multicolumn{2}{c}{interaction} &
         \multicolumn{2}{c}{$\epsilon$} & $E_\Delta$ \\
         \cline{3-4}\cline{5-6}\rule{0pt}{2ex}%
         &&type 1&type 2& Categor. & Hyper.&\\
         \hline
    \rule{0pt}{2ex}%
         Zachary's karate club~\cite{zachary_1977_InformationFlowModel}    & 34  & 11  & 67  &         0.13  & \textbf{0.11} & 0    \\
         Crimes~\cite{decker_1991_StLouisHomicide}                         & 202 & 57  & 209 &         0.11  & \textbf{0.05} & 0    \\
         Sexual contacts~\cite{rocha_2011_SimulatedEpidemicsEmpirical}     & 159 & 47  & 108 &         0.11  & \textbf{0.05} & 0    \\
         Plant-pollinator~\cite{kato_1990_InsectflowerRelationshipPrimary} & 57  & 51  & 128 &         0.11  &         0.11  & 0.80 \\
         Languages~\cite{kunegis_2013_KONECTKoblenzNetwork}                & 150 & 30  & 242 &         0.07  & \textbf{0.03} & 0    \\[5pt]
         Hypergraph SBM~\cite{paul_2018_HigherOrderSpectralClustering}     & 100 & 60  & 76  &         0.44  & \textbf{0.03} & 0.20 \\
         Triangle-edge CM~\cite{miller_2009_PercolationEpidemicsRandom}    & 100 & 107 & 89  &         0.52  & \textbf{0.09} & 0.32 \\
         $\beta$-model~\cite{stasi_2014_BetaModelsRandom}                  & 100 & 61  & 56  &         0.51  & \textbf{0.33} & 0.70 \\[5pt]
         Best-case                                                         & 100 & 92  & 93  &         0.38  & \textbf{0.01} & 0    \\
         Worst-case                                                        & 100 & 100 & 100 & \textbf{0.36} &         0.50  & 1    \\
    \hline
    \hline
    \end{tabular}
\end{table*}
}

\subsection{Case study: Zachary's Karate Club}

We first illustrate the framework with a simple case study based on Zachary's Karate Club~\cite{zachary_1977_InformationFlowModel}.
Our goal will be to recover the latent structure of this system, encoded as a hypergraph $H$, given synthetic data $X$ generated with the likelihood of Eq.~\eqref{eq:likelihood} and  $\mu_0=0.01, \mu_1=20$ and $\mu_2=30$.
These make it fairly easy to discern non-interacting pairs but lead to some overlap between the two other types of interactions, which will allow us to highlight the influence of higher-order interactions on the accuracy on the inference (see Fig.~\ref{fig:mixture_dist} which illustrates the distribution of pairwise measurement for this choice of parameters).
The structure of the original Karate Club only contains dyadic observations which makes for an uninteresting test of our method, so we add the 3-edges that are found by a separate hypergraph inference technique~\cite{young_2021_HypergraphReconstructionNetwork}.
(We break down any hyperedge larger than 3 vertices into multiple 3-edges.)
We show the original graph and associated hypergraph in Figs.~\ref{fig:zachary_graphs}a~and~\ref{fig:zachary_graphs}b---we use the latter throughout our case study.

With this hypergraph structure fixed, we generate a synthetic dataset $X$ and approximate the posterior distribution using samples generated with the Monte Carlo Markov chain (MCMC) algorithms described in Appendix~\ref{annex:gibbs}.
Using these samples, we then calculate two estimators of the structure: the maximum \posterior{} (MAP) estimator
\begin{align}
    \hat{\mathcal{S}}_\text{MAP} = \argmax\limits_{\mathcal{S}} P(\mathcal{S}|X),
\end{align}
corresponding to the latent structure that maximizes the posterior distribution, and the edge-wise estimator $\hat{\mathcal{S}}_\text{EW}$ that only contains the weak/strong edges or 2-edges/3-edges with a marginal posterior probability above 0.5, e.g., for the hypergraph model
\begin{align}
    \hat{\mathcal{S}}_\text{EW} =  \{ e | \; e \in E\cup T, P(e|X) > 0.5\},
\end{align}
where $P(e|X)$ is the marginal probability that edge $e$ is present.
We complement these structural estimators with an estimator of the type of each pairwise interaction, the maximum marginal estimator
\begin{subequations}
\begin{align}
    \hat{\mathcal{S}}_\text{MM} =  \{\hat{\ell}_{ij}\, | \; i,j \in V\},
\end{align}
where
\begin{align}
    \hat{\ell}_{ij} = \argmax\limits_{\ell_{ij} \in \{0,1,2\}} P(\ell_{ij}|X)
    \label{eq:ell_hat}
\end{align}
\end{subequations}
is the most likely type of interaction type for vertices $i$ and $j$ (ties are broken by choosing a type at random).

Figures~\ref{fig:zachary_graphs}c~and~\ref{fig:zachary_graphs}d show $\hat{\mathcal{S}}_\text{MAP}$ and  $\hat{\mathcal{S}}_\text{EW}$ for both models given one realization of the data $X$. In both cases, we see that our inference framework reconstructs the original structure quite accurately. Interestingly, we see that both estimators missed a few 3-edges.  While some of them are genuine errors, quite a few missing 3-edges are simply hidden and thus unrecoverable (as defined in section~\ref{ssec:s_models}).

Thankfully, these missing hidden 3-edges have little impact on the accuracy of our framework when it comes to predicting the interaction label $\ell_{ij}$. To show this, Figs.~\ref{fig:zachary_graphs}e~and~\ref{fig:zachary_graphs}f also display confusion matrices, a generalization of statistical errors (type I and type II errors) for multiple classes. In the confusion matrix, the element $c_{rs}$ is the number of times a pairwise interaction of type $\ell_{ij}=r$ has been predicted as $\hat \ell_{ij}=s$ by the maximum marginal estimator $\hat{\mathcal{S}}_\text{MM}$. Hence, a perfect reconstruction corresponds to a diagonal matrix.
The major difference between both confusion matrices is that the categorical edges graph model uses weak edges and strong edges somewhat interchangeably, which results in classification errors in both ways. In contrast, the hypergraph model has no false positive 3-edges. This is due to the restrictive nature of 3-edges: each type-2 pairwise interaction must be associated with at least two other type-2 pairwise interactions (as long as the 3-edge is not hidden). As a result, our framework will err on a more conservative side when  assigning larger hyperedges: The framework will assign $\ell_{ij} = 1$ unless there is sufficient evidence in the neighborhood of vertices $i$ and $j$ that supports a 3-edge. This additional neighborhood information is what allows the hypergraph model to have a smaller sum of off-diagonal elements in the confusion matrix, meaning that it more accurately retrieves the interaction types.

\begin{figure}
    \centering
    \includegraphics[width=\linewidth]{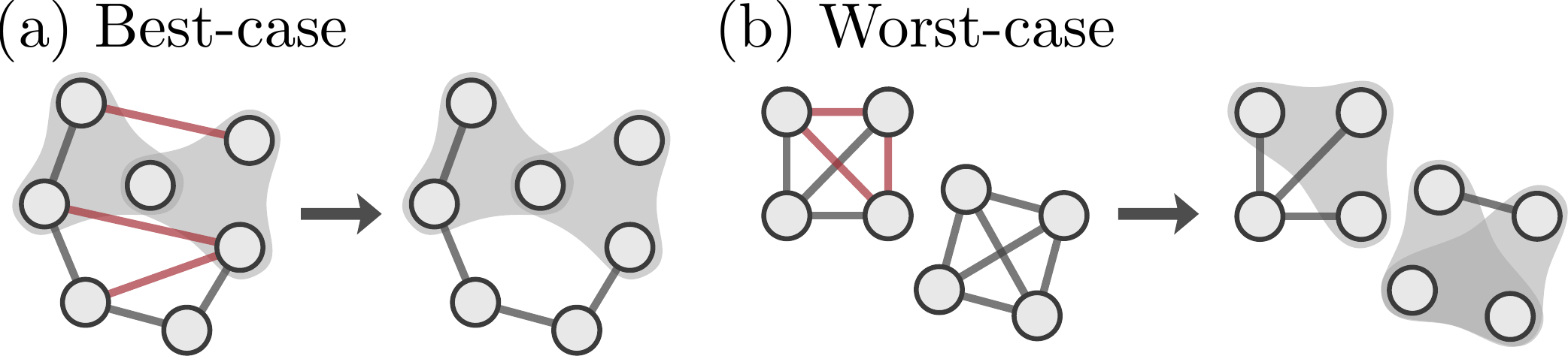}
    \caption{Illustration of the generation of the best-case and worst-case hypergraphs.
    (a) The best-case hypergraphs are obtained by first generating a random hypergraph using Eq.~\eqref{eq:phg_prior} and then removing any 2-edge that creates triangle when projecting the hypergraph onto the pairwise interactions.
    (b) The worst-case hypergraphs are generated from a graph with cliques of 2-edges and in which each triangle can be promoted to a 3-edge with a given probability.}
    \label{fig:hypergraph_gen}
\end{figure}

\subsection{Expanded dataset}

Next, we study the performance of our inference framework on a broader collection of synthetic and empirical hypergraphs.
The empirical datasets are a crime network~\cite{decker_1991_StLouisHomicide}, a network of sexual contacts~\cite{rocha_2011_SimulatedEpidemicsEmpirical}, a plant--pollinator network~\cite{kato_1990_InsectflowerRelationshipPrimary} and a network of spoken languages~\cite{kunegis_2013_KONECTKoblenzNetwork}. The original datasets are all bipartite, so we again adapt them for the experiment by interpreting one of the two vertex types as hyperedges: individuals are vertices and crimes are hyperedges, sex workers are vertices and hyperedges are their clients, pollinators are the vertices and the plants they pollinate are hyperedges, vertices are countries and hyperedges are languages spoken. We ignore hyperedges with more than five vertices to keep a sufficient number of 2-edges in the hypergraph, and we also remove any isolated vertex.
We also re-use the hypergraph derived from Zachary’s karate above.

We complement these empirical datasets with hypergraphs generated with the three computer models, namely (i) the superimposed stochastic block model~\cite{paul_2018_HigherOrderSpectralClustering} (two unequal communities of 30 and 70 vertices with connection probabilities of $q_{11}=0.05$, $q_{12}=q_{21}=0.001$ and $q_{22}=0.02$ for 2-edges, and of $p_1=0.005$ and $p_2=0.0001$ for 3-edges inside communities and $p_\text{out}=0.00001$ outside communities),  (ii) a triangle-edge configuration model of 100 vertices~\cite{miller_2009_PercolationEpidemicsRandom} (with degrees drawn from independent geometric distributions of mean $2$ and $3$ for 2-edges and 3-edges, respectively), and (iii) the $\beta$-model for layered hypergraphs~\cite{stasi_2014_BetaModelsRandom} (with vertex propensities of 2-edges and 3-edges drawn from normal distributions of averages $-4.5$ and $-5$ and of standard deviations $2.5$ and $2$, respectively).

As before, we generate a series of synthetic observations with the likelihood in Eq.~\eqref{eq:likelihood} and $\mu=(\mu_0=0.01, \mu_1=40, \mu_2=50)$, and then sample the posterior distribution to compute the confusion matrices of both models. We summarize our results using the fraction of misclassified type-1 and type-2 interactions, a quantity we call the \emph{relative reconstruction error}
\begin{align}
    \epsilon = \frac{c_{10} + c_{12} + c_{20} + c_{21}}{c_{10} + c_{11} + c_{12} + c_{20} + c_{21} + c_{22}},
\end{align}
where $c_{rs}$ are the entries of the confusion matrix.
The results are reported in Table~\ref{tab:real_data_results} where we see that the hypergraph model performs at least as well as the categorical-edge model.
The following section explores the factors influencing the performance of the hyperedge model.

\subsection{When are the hyperedges most relevant}
\label{ssec:synthetic_inference}

To gain better insights on the factors influencing the performance of the hyperedge model, we consider two extreme cases: a ``worst-case hypergraph'' and a ``best-case hypergraph''.

In the best-case hypergraphs, groups of 3 vertices can only be connected by a 3-edge. This means that vertices $(i,j,k)$ can form a triangle in projected pairwise interaction only if $\ell_{ij}=\ell_{ik}=\ell_{jk}=2$. As a result, there is no ambiguity on whether or not triangles are a mix of 2-edges and projected 3-edges, and 3-edges can be distinguished from triangles of non-interacting pairs since they have greater pairwise measurements.
This effectively makes the neighborhood of any pair of vertices very informative on its type of interaction. We generate such hypergraphs by removing the 2-edges that do not respect the imposed constraint from a hypergraph generated with the prior~\eqref{eq:phg_prior} (see Fig.~\ref{fig:hypergraph_gen}).

The worst-case hypergraphs only contain 2-edges if they form a triangle in the projection. In other words, $\ell_{ij}=1$ is only possible if there exists another vertex $k$ such that $\ell_{ik}\ell_{jk}>0$. As a result, there is no longer a difference in the observations between a 3-edge and a triangle comprised of a mixture of 2-edges and projected 3-edges; the neighborhood of a pairwise observation is uninformative. To produce these worst-case hypergraphs, we generate graphs with cliques of 2-edges where each triangle is promoted randomly to a 3-edge (see Fig.~\ref{fig:hypergraph_gen}).

To check whether a given hypergraph resembles the best-case or the worst-case, we compute the proportion of 2-edges inside triangles
\begin{align}
    E_\Delta &= \frac{1}{h_1} \sum_{(i,j) \in E} \mathds{1}[(i,j) \in \Delta].
\end{align}
The closer $E_\Delta$ is to $0$, the closer the hypergraph is to a best-case hypergraph, and the closer the $E_\Delta$ is to $1$, the closer the hypergraph is to a worst-case hypergraph.

Revisiting Table~\ref{tab:real_data_results}, we see that $E_\Delta$ is related to the error $\epsilon$ and that errors for each hypergraph range between the best-case and the worst-case. However, the proportions $\{\rho_k\}$ also play a role in $\epsilon$: when a type of interaction is being observed at a similar rate to another, models will most likely favor the type with the largest proportion as it leads to a better fit.

Table~\ref{tab:real_data_results} also shows that empirical hypergraphs are generally closer to a best-case hypergraph than to a worst-case. This is due to the sparsity of interactions of empirical complex systems: we expect that most 2-edges are not part of projected triangles. For that reason, the hypergraph model works better than the categorical edges graph model for the majority of systems.
And when the hypergraph model errs, both models tend to err as confirmed by the last two lines of Table~\ref{tab:real_data_results}.

\subsection{Impact of data means}
\begin{figure}
    \subfloat[\label{sfig:best_confusion_mu1}]{%
        \includegraphics[width=.47\columnwidth]{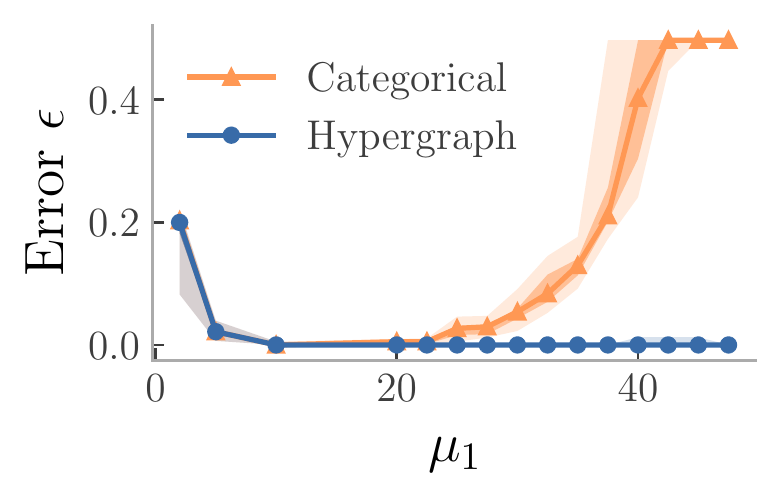}
    }\hfill
    \subfloat[\label{sfig:best_entropy_mu1}]{%
        \includegraphics[width=.47\columnwidth]{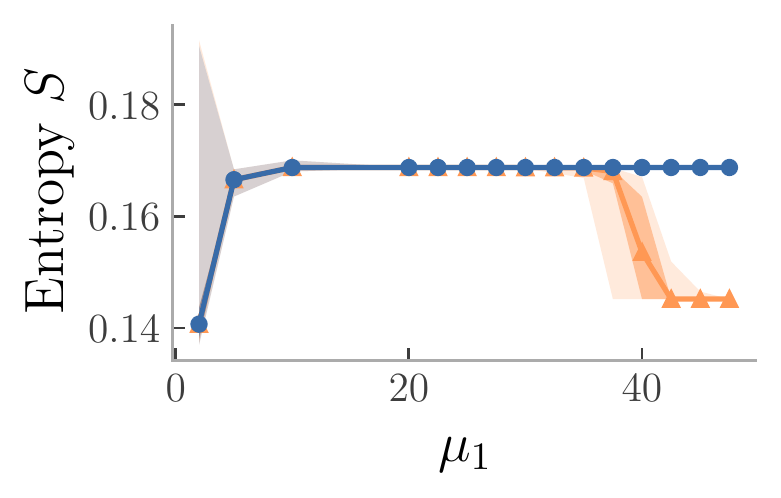}
    } \newline
    \subfloat[\label{sfig:best_residuals_mu1}]{%
    \includegraphics[width=\linewidth]{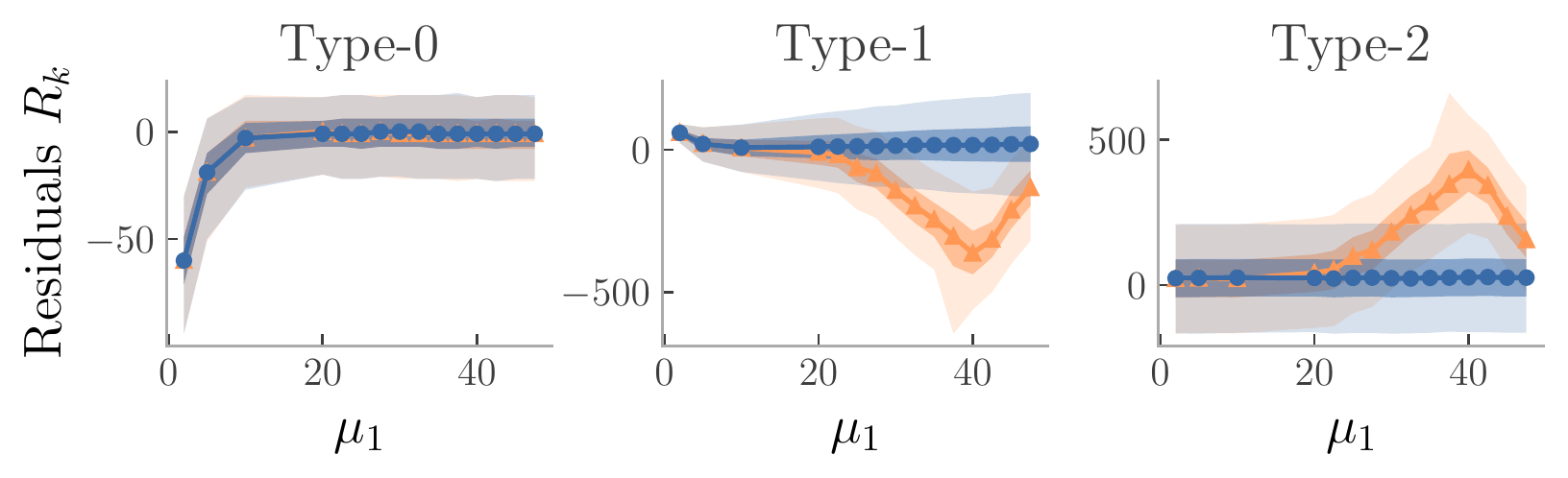}
    }
    \caption{
    Impact of the measurement rate ($\mu_1$) of type-1 interactions on the reconstruction of a best-case hypergraph.
    (a) Relative reconstruction error $\epsilon$. (b) Entropy $S$. (c) Sums of residuals $R_k$. The observations were generated with $\mu_0=0.01$, $\mu_2=50$ and various $\mu_1$ using the hypergraph model (blue) and the categorical edges graph model (orange). The hypergraph model displays (a) a smaller misclassification error (b) a larger entropy and (c) lower residuals than the categorical edges graph model, which indicates a better reconstruction.
    Symbols represent the median, light colored shadings are percentiles 2.5 and 97.5 and dark colored shadings are percentiles 25 and 75 of the metrics for 100 synthetic observations.
    Residuals were evaluated using 200 predictive observation matrices and the best-case hypergraph was generated using $p=0.00017$ and $q=0.019$.
    }
    \label{fig:best_mu1}
\end{figure}

\begin{figure}
    \subfloat[\label{sfig:worst_confusion_mu1}]{%
        \includegraphics[width=.47\columnwidth]{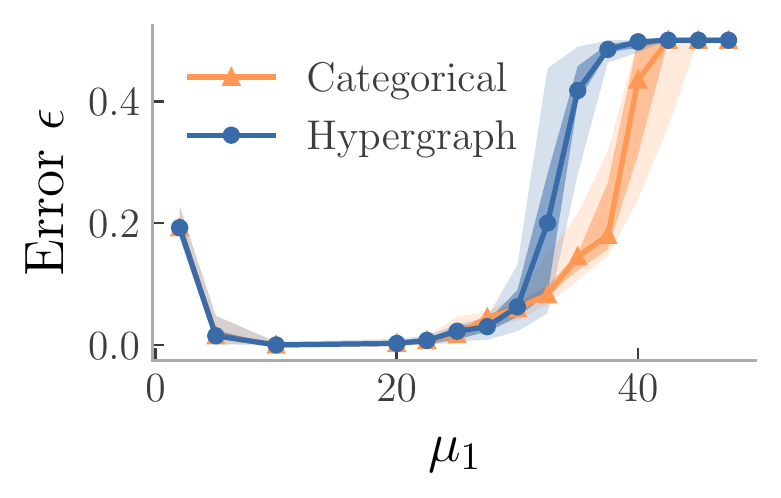}
    }\hfill
    \subfloat[\label{sfig:worst_entropy_mu1}]{%
        \includegraphics[width=.47\columnwidth]{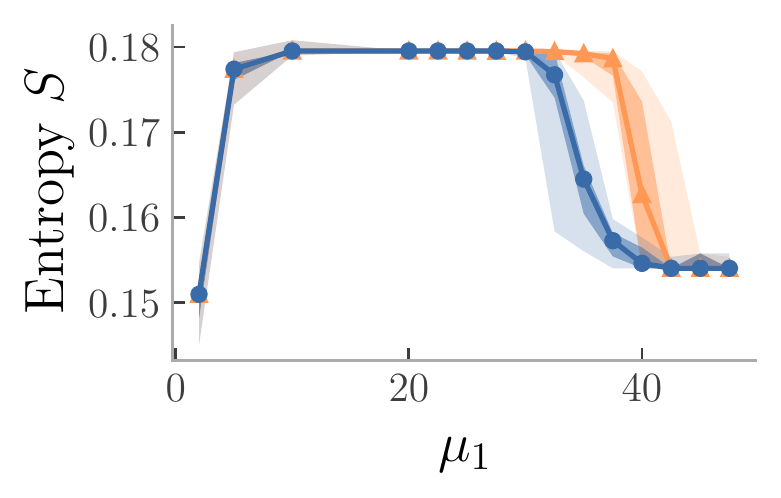}
    } \newline
    \subfloat[\label{sfig:worst_residuals_mu1}]{%
    \includegraphics[width=\linewidth]{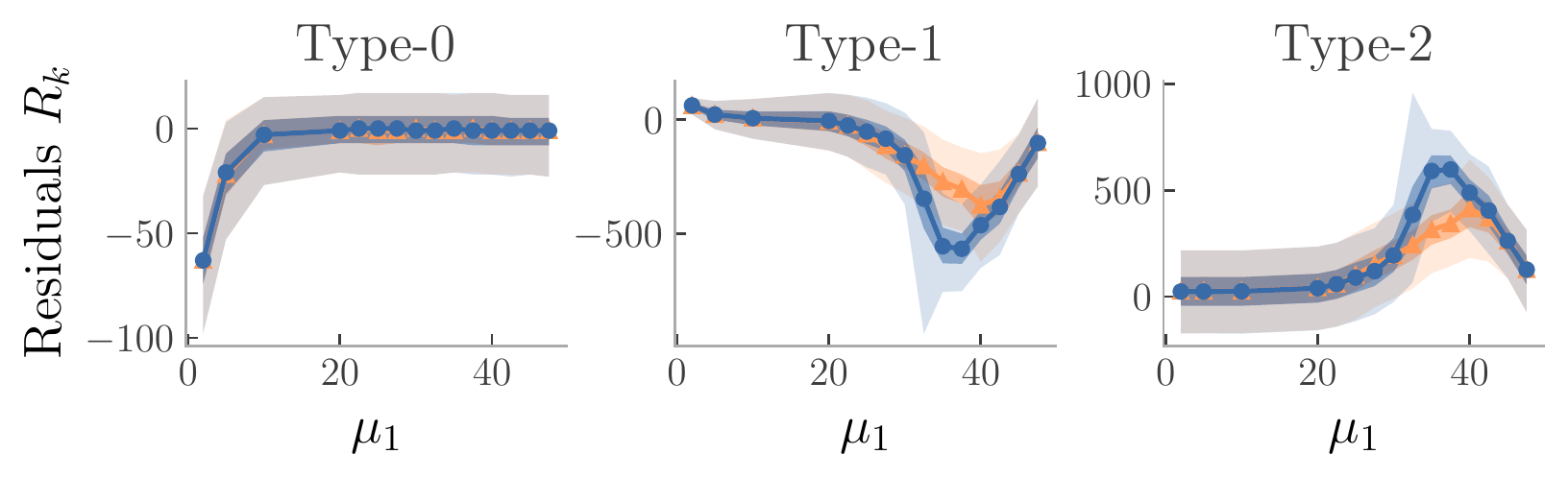}
    }
    \caption{
    Impact of the measurement rate ($\mu_1$) of type-1 interactions on the reconstruction of a worst-case hypergraph (see Fig.~\ref{fig:best_mu1} for details).
    While the categorical edges graph has a similar performance to the best-case hypergraph (Fig.~\ref{fig:best_mu1}), the hypergraph model cannot distinguish 3-edges from 2-edges with triangles, which results in a worse reconstruction. This is seen with (a) a larger misclassification error (b) a smaller entropy and (c) larger residuals.
    The worst-case hypergraph was constructed from 20 5-cliques in which triangles were promoted to 3-edges with probability $0.19$.
    }
    \label{fig:worst_mu1}
\end{figure}

To complete our analysis, we study the impact of the parameters $\mu$ on the reconstruction by varying $\mu_1$ while keeping $\mu_0=0.05$ and $\mu_2=50$ fixed, for the two families of extreme hypergraphs described above (with $n=100$ vertices).
Doing so allows us to identify the regimes in which the hypergraph model displays a better performance. In addition to the relative reconstruction error $\epsilon$, we also consider two additional summary statistics: the entropy $S$ of the label distribution, and the sums of residuals $R_k$.

We define the entropy of the label distribution as
\begin{subequations}
\begin{align}
    S & = -\sum_{k=0}^2 \rho_k \log_3 \rho_k,
    \intertext{where}
    \rho_k & = \frac{c_{0k} + c_{1k} + c_{2k}}{\binom{n}{2}}
\end{align}
\end{subequations}
is the proportion of pairs predicted as type $k$. This statistic measures the effective number of interactions predicted by the models: it is $0$ if only one type of interaction exists and it is $1$ if $\rho_0=\rho_1=\rho_2=\frac{1}{3}$. Because the empirical datasets we consider are sparse, most pairs of vertices do not interact, meaning that $S$ is small. Nevertheless, comparing entropy values allows us to detect when a model completely ignores a type of interaction.

The sums of residuals $R_k$ are defined as
\begin{align}
    R_k &= \sum_{i<j} (x_{ij} - \tilde x_{ij}) \delta_{k,\ell_{ij}},
\end{align}
where $\tilde X = [\tilde x_{ij}]_{i,j=1,\dots, n}$ is an observation matrix generated synthetically from the posterior-predictive distribution~\cite{gelman_1996_PosteriorPredictiveAssessment,young_2021_BayesianInferenceNetwork}.
For each sample point $\mathcal{\tilde S}, \tilde \mu \sim P(\mathcal{S},\mu|X)$, we generate predictive matrices $\tilde{X}$ from the likelihood~\eqref{eq:likelihood}.
This is known as a form of \textit{posterior--predictive check}, and it quantifies the goodness of fit of a model by checking that the fitted model can adequately reproduce the original data. The statistics $R_k$ will reveal biases in the fitted model,  with $R_k\approx 0$ only when  the predicted pairwise observations $\tilde x_{ij}$ are on average equal to the pairwise observations $x_{ij}$ for the interactions of type $k$.

Figures~\ref{fig:best_mu1} and~\ref{fig:worst_mu1} show that the relative reconstruction error generally increases as $\mu_1$ approaches $\mu_0$ or $\mu_2$. This behavior is expected because there is a greater overlap between the corresponding Poisson distributions in the observations $X$. When this overlap is large, interaction types are represented similarly in the observations $X$, which makes them difficult to infer. Figures~\ref{fig:best_mu1} and~\ref{fig:worst_mu1} also show that the entropy generally decreases and stabilizes to a lower plateau as $\mu_1$ approaches $\mu_2$. This is due to a similar phenomenon: with the increasing overlap, models favor one type of interaction over the other to the point where one type of interaction disappears. Once the interaction types have ``merged'', the entropy remains constant.

For the best-case hypergraph, we clearly see in Fig.~\ref{fig:best_mu1} that the hypergraph model overall outperforms the categorical edges graph model. Figure~\ref{sfig:best_confusion_mu1} shows that the hypergraph model makes very little classification errors for all sets of parameters. This translates to a higher entropy, as seen in Fig.~\ref{sfig:best_entropy_mu1}, and to a smaller predictive bias in Fig.~\ref{sfig:best_residuals_mu1}. We conclude that the worse performance observed for the categorical edges graph model is explained by weak and strong edges ending up being interchangeable because of their pairwise nature. Without the information from the neighborhood that 3-edges imply, the interaction type of a pair $\ell_{ij}$ must be deduced from its observation $x_{ij}$ alone.

For the worst-case hypergraph, Fig.~\ref{fig:worst_mu1} illustrates that the categorical edges graph model slightly outperforms the hypergraph model. We believe this is due to the prior distribution of the 3-edge probability $p$: because there are $\binom{n}{3}$ possible 3-edges compared to $\binom{n}{2}$ possible 2-edges, there is a much larger number of 3-edges than strong edges for the same probability. In this worst-case setting, 3-edges are almost indistinguishable from 2-edges since triangles are mixture of 2-edges and projected 3-edges. Thus, there is no improvement brought by the hypergraph model, which suggest that this hypergraph representation is not appropriate.

\section{Conclusion}

Mounting evidence collected in recent years support that the behavior of many complex systems require taking into account high-order interactions. However, many of the tools of this rapidly expanding field have yet to find practical applications still as measurements of higher-order systems remains challenging to this day

We presented a minimal Bayesian inference framework that makes progress in this direction, by reconstructing hypergraphs from noisy observations of their pairwise projection.  Using synthetic and empirical datasets, we illustrated the impact that taking into account high-order interactions has on the accuracy of the reconstruction.  Notably, we identified the regimes where high-order interactions yield fewer reconstruction errors, due to the fact that hyperedges require the use of local information contained in the neighborhood of vertices.

Although the inference framework introduced here is fairly general, we illustrated it using simple data and hypergraph models to avoid obfuscating its presentation unnecessarily.  Thus, future work should be done to apply our framework to hypergraphs with hyperedges larger than 3-edges, and to non-Poissonian data models. Doing so will require to treat carefully the way higher-order interactions are assumed to be encoded in the pairwise observation data; as we have shown, hidden hyperedges can hinder high quality reconstruction. A possible solution worth investigating involves the use of simplicial complexes, a more restricted higher-order structure in which a hyperedge of size $k$ implies every hyperedge of size $k-1$.  Yet, how to connect pairwise interactions to such higher-order interactions remains an open question and is a testament to the bright future Bayesian inference of higher-order interactions has over the coming years.

\section*{Acknowledgments}
We thank Charles Murphy for helpful comments and suggestions on preliminary versions of this work.
This work was supported by the Conseil de recherches en sciences naturelles et en g\'enie du Canada (SL, AA), the Sentinelle Nord program of Universit\'e Laval (SL, AA), funded by the Fonds d’excellence en recherche Apog\'ee Canada, and the James S. McDonnell Foundation (JGY).
We acknowledge Calcul Qu\'ebec and Alliance de recherche numérique du Canada for their technical support and computing infrastructures.

\medskip


%

\clearpage
\appendix

\section{Prior distributions}
\label{annex:priors}

We use the conjugate priors for each parameter in the model, which corresponds to Beta distributions
\begin{subequations}
\label{eq:prob_priors}
\begin{align}
    q_1 & \sim \text{Beta}(\xi, \zeta) \\
    q_2 & \sim \text{Beta}(\xi, \zeta) \\
    p   & \sim \text{Beta}(\xi, \zeta) \\
    q   & \sim \text{Beta}(\xi, \zeta).
\end{align}
\end{subequations}
In all experiments we set $\xi=1.1$ and $\zeta=5$ which encourages sparsity while discouraging the complete removal of a interaction types (a null probability).

As discussed in the main text, we address the potential label switching problem of edge types by imposing an order for the parameters $\mu = (\mu_0, \mu_1, \mu_2)$, which can be viewed as a prior on these parameters~\cite{betancourt_2017_IdentifyingBayesianMixture}. For the categorical-edge model, we impose a total ordering $\mu_0<\mu_1<\mu_2$ while the correlations produced by the triangles of the hypergraph model allow us to only assume the partial order $\mu_0<\mu_1$ and $\mu_0<\mu_2$ under the assumption that the difference of hyperedge size is sufficient to break symmetries. These considerations translate into the following conjugate distributions for the categorical-edges model
\begin{subequations}
\label{eq:pes_param_prior}
\begin{align}
    \mu_0 &\sim \text{Gamma}(\alpha_0, \beta_0),\\
    \mu_1|\mu_0 &\sim \TruncGamma_{(\mu_0, \infty)}(\alpha_1, \beta_1) \label{eq:pes_ordered_mu_priors}, \\
    \mu_2|\mu_1 &\sim \TruncGamma_{(\mu_1, \infty)}(\alpha_2, \beta_2),
\end{align}
\end{subequations}
and for the hypergraph model we have
\begin{subequations}
\label{eq:phg_param_prior}
\begin{align}
    \mu_0 &\sim \text{Gamma}(\alpha_0, \beta_0),\\
    \mu_1|\mu_0 &\sim \TruncGamma_{(\mu_0, \infty)}(\alpha_1, \beta_1), \label{eq:phg_ordered_mu_priors} \\
    \mu_2|\mu_0 &\sim \TruncGamma_{(\mu_0, \infty)}(\alpha_2, \beta_2).
\end{align}
\end{subequations}
We use the following probability density functions
for $x\sim \text{Gamma}(\alpha, \beta)$ and $y\sim \text{TruncGamma}_{(c,d)}(\alpha, \beta)$:
\begin{align}
    f(x) &= \frac{1}{\Gamma(\alpha)} x^{\alpha-1}e^{-\beta x} \mathds{1}_{(a, b)}(x), \\
    g(y) &= \frac{1}{\gamma(d, \alpha)-\gamma(c, \alpha)} \frac{1}{\Gamma(\alpha)} y^{\alpha-1}e^{-\beta y} \mathds{1}_{(c, d)}(y),
\end{align}
where $\gamma$ is the lower incomplete gamma function and $\mathds{1}$ is the indicator function.
In all our numerical experiments, we set the priors $\alpha_0 = \alpha_1 = \alpha_2 = 1.05$ and $\beta_0 = \beta_1 = \beta_2 = 0.5$.
In another inference setting, these should be adjusted to reflect prior knowledge about the dataset X.

\section{Sampling algorithms}
\label{annex:gibbs}

We use a Gibbs sampler to sample the joint posterior distribution $P(\mathcal{S}, \theta |X)$, where $\theta = \{\mu, \phi\}$. This class of algorithms allows us to sample from arbitrary joint distributions by sampling from each of its conditional distributions in alternance, here the parameter distribution $P(\theta|\mathcal{S},X)$ and the structural distribution $P(\mathcal{S}| X, \theta)$. In what follows, we derive these sampling distributions and determine algorithms that generate samples from them.

\subsection{Sampling the parameters}
\label{annex:parameters_sampling}

We break down the sampling of the parameters $\theta$ in sequential sampling steps for each of the individual parameters, meaning that when sampling from $P(\theta|\mathcal{S}, X)$, each parameter is conditionally independent to the others. This marginal distribution is noted $P(\theta^*|\theta_{-\theta^*}, \mathcal{S}, X)$ where $\theta_{-\theta^*}$ represents all parameters excluding $\theta^*$. Using Bayes formula, one can see that this distribution is proportional to the posterior distribution
\begin{align}
    P(\theta^*| \theta_{-\theta^*}, \mathcal{S}, X) = \frac{P(\mathcal{S}, \theta| X)}{P(\mathcal{S}|X)} \propto P(\mathcal{S}, \theta| X).
    \label{eq:theta_full_conditional}
\end{align}
Using Eqs.~\eqref{eq:phg_prior}, \eqref{eq:pes_prior} and \eqref{eq:prob_priors}, we directly find that
\begin{subequations}
\begin{align}
    q_1|\theta_{-q_1}, G, X & \sim \text{Beta} (m_1+\xi, {\scriptstyle\binom{n}{2}} - m_1 - m_2 + \zeta) \\
    q_2|\theta_{-q_2}, G, X & \sim \text{Beta} (m_2+\xi, {\scriptstyle\binom{n}{2}} - m_2       + \zeta) \\
    q|\theta_{-q}, H, X     & \sim \text{Beta} (h_1+\xi, {\scriptstyle\binom{n}{2}} - h_1       + \zeta) \\
    p|\theta_{-p}, H, X     & \sim \text{Beta} (h_2+\xi, {\scriptstyle\binom{n}{3}} - h_2       + \zeta),
\end{align}
\end{subequations}
which are all beta distributions. A random variable $z\sim \text{Beta}(a, b)$ can be sampled rapidly with standard univariate sampling methods available in most statistical
software packages, for example as $z=x/(x+y)$ where $x\sim\text{Gamma}(a)$ and $y\sim\text{Gamma}(b)$~\cite{ahrens_1974_ComputerMethodsSampling}.

To sample the parameters $\mu = (\mu_0, \mu_1, \mu_2)$, we rearrange the product inside Eq.~\eqref{eq:likelihood} as
\begin{align}
    P(X|\mathcal{S}, \theta) &= \prod_{i<j} \qty( \frac{1}{x_{ij}!} ) \prod_{k=0}^2 \mu_k^{X^{(k)}} e^{-\mu_k L^{(k)}}
\end{align}
where
\begin{align}
  X^{(k)} & = \sum_{i<j} x_{ij} \delta_{k,\ell_{ij}} \label{eq:x_pow_k}\\
  L^{(k)} & = \sum_{i<j}        \delta_{k,\ell_{ij}} \label{eq:l_pow_k}
\end{align}
are, respectively, the sum of observations with label $k$ and the number of pairs with label $k$, and where $\delta$ being the Kronecker delta. Combining Eqs.~\eqref{eq:pes_prior}~and~\eqref{eq:theta_full_conditional} yields for the categorical edges graph model
\begin{subequations}
\begin{align}
    \mu_0|\theta_{-\mu_0},G,X & \sim \TruncGamma_{(0,    \mu_1)}(X^{(0)}\!\!+\!\alpha_0, L^{(0)}\!+\!\beta_0) \\
    \mu_1|\theta_{-\mu_1},G,X & \sim \TruncGamma_{(\mu_0,\mu_2)} (X^{(1)}\!\!+\!\alpha_1, L^{(1)}\!+\!\beta_1) \\
    \mu_2|\theta_{-\mu_2},G,X & \sim \TruncGamma_{(\mu_1,\infty)}(X^{(2)}\!\!+\!\alpha_2, L^{(2)}\!+\!\beta_2).
\end{align}
\end{subequations}
Combining Eqs.~\eqref{eq:phg_prior}~and~\eqref{eq:theta_full_conditional} yields for the hypergraph model
\begin{subequations}
\begin{align}
    \mu_0|\theta_{-\mu_0},H,X & \sim \TruncGamma_{(0,    \mu_-)} (X^{(0)}\!\!+\!\alpha_0, L^{(0)}\!+\!\beta_0) \\
    \mu_1|\theta_{-\mu_1},H,X & \sim \TruncGamma_{(\mu_0,\infty)}(X^{(1)}\!\!+\!\alpha_1, L^{(1)}\!+\!\beta_1) \\
    \mu_2|\theta_{-\mu_2},H,X & \sim \TruncGamma_{(\mu_0,\infty)}(X^{(2)}\!\!+\!\alpha_2, L^{(2)}\!+\!\beta_2)
\end{align}
\end{subequations}
where $\mu_- = \min\{\mu_1,\mu_2\}$.

Since this step is revisited often by our algorithm, we combine three sampling methods to ensure rapid and accurate sampling in all cases~\cite{gallagher_2021_ClarifiedTypologyCoreperiphery}: rejection sampling using a gamma distribution if the rejection probability is low, a more costly inverse transform sampling using incomplete gamma inverse function, and rejection sampling with an adjusted ``linear distribution'' if all other methods fail. The main interest in using the linear distribution is that it provides a good approximation of the density for small intervals. The inverse transform sampling often works, but can suffer from numerical instabilities especially for small truncation intervals.

We define the linear probability density function as
\begin{align}
    f(x) = \frac{1+cx}{2}, \quad x,c \in[-1,1]
\end{align}
where $c$ is the slope. A sample from this distribution is obtained using its inverse cumulative distribution function
\begin{align}
    \text{CDF}^{-1}(u) = \frac{\sqrt{c^2-2c+4cu+1}-1}{c}
\end{align}
where $u$ is a continuous random variable uniformly distributed on $[0,1]$.

In the rejection sampling algorithm, the support of this distribution is adjusted to match the truncated gamma distribution and $c$ is the slope of a line connecting the truncated gamma density evaluated at the lower bound to the density evaluated at the upper bound.

\subsection{Sampling graphs with categorical edges}
\label{annex:graph_sampling}

The distribution used to sample the categorical edges graph model is derived by following a similar reasoning as for Eq.~\eqref{eq:theta_full_conditional}. We first observe that
\begin{align}
    P(\mathcal{S}|\theta, X) = \frac{P(\mathcal{S}, \theta| X)}{P(\theta|X)} \propto P(\mathcal{S}, \theta| X).
    \label{eq:s_full_conditional}
\end{align}
Combining this expression with Eqs.~\eqref{eq:pes_posterior}~and~\eqref{eq:prob_priors} yields
\begin{multline}
    P(G|\theta, X) \propto  q_1^{m_1+\xi-1} (1-q_1)^{\binom{n}{2}-m_1-m_2+\zeta-1} \\
    \times q_2^{m_2+\xi-1} (1-q_2)^{\binom{n}{2} - m_2+\zeta-1} \\
    \times \prod_{i<j} \frac{ (\mu_{\ell_{ij}})^{x_{ij}} }{x_{ij}!} e^{-\mu_{\ell_{ij}}}.
\end{multline}
The edge labels $\ell_{ij}$ induce complicated interactions between the parameters, so we turn to a Metropolis-Hastings (MH) algorithm to generate samples from this distribution as it does not appear to correspond to a well known closed-form distribution.

The MH algorithm is initialized at the ground truth hypergraph projection and the ground truth parameters except in Table~\ref{tab:real_data_results} where it is initialized at a graph with no strong edges and weak edges wherever $x_{ij} > 0$ and at parameters $\mu$ and $\phi_G$ set to the maximum likelihood estimator obtained from a Poisson mixture model.
At each iteration, we propose to increment a interaction type with probability $\eta$ and to decrement a interaction type with probability $1-\eta$. We use $\eta=0.5$ in our numerical simulations.

If the algorithm reaches a point where the graph is fully connected with strong edges (or empty), than we propose to decrement (or increment) a type with probability 1.
The pair $(i,j)$ whose type is to be decremented is chosen uniformly among all pairs whose type is not zero. The pair $(i,j)$ whose type is to be incremented is chosen proportionally to the weight
\begin{align}
    w_{ij} = \begin{cases}
    x_{ij}+1 & \text{if }\ell_{ij}<2\\
    0 & \text{otherwise.}
    \label{eq:edge_weight}
    \end{cases}
\end{align}
The proposal probability of a new graph $G^*$ conditioned on the current graph $G$ is
\begin{align}
    Q(G^*|G, X) = a \frac{\eta w_{ij}}{\sum_{i<j} w_{ij}} + (1-a) \frac{1-\eta}{m_1+m_2}
    \label{eq:mh_pes}
\end{align}
where $a=1$ if the label is to be incremented and $a=0$ if it is to be decremented. Finally, the proposal is accepted with probability
\begin{align}
    \alpha(G^*|G) = \min\qty(1,\ \frac{P(G^*, \theta|X)Q(G|G^*, X)}{P(G,\theta|X)Q(G^*|G, X)})
\end{align}
where $Q(G|G^*, X)$ is the probability of reverting the proposed move.

\subsection{Sampling hypergraphs}
\label{annex:hypergraph_sampling}

Combining Eqs.~\eqref{eq:phg_posterior}, \eqref{eq:prob_priors} and~\eqref{eq:s_full_conditional}, we find
\begin{multline}
    P(H|\theta, X) \propto \frac{P(\theta)}{P(X)} q^{h_1+\xi-1} (1-q)^{\binom{n}{2}-h_1+\zeta-1} p^{h_2+\xi-1}  \\
    \times (1-p)^{\binom{n}{3}-h_2+\zeta-1} \prod_{i<j} \frac{ (\mu_{\ell_{ij}})^{x_{ij}} }{x_{ij}!} e^{-\mu_{\ell_{ij}}},
    \label{eq:phg_h_full_conditional}
\end{multline}
which, again, is not a standard distribution. Hence we use a MH algorithm to generate samples of it in a similar fashion as for the categorical edges graph model.

\begin{figure}[t]
    \subfloat[\label{sfig:best_confusion_mu2}]{%
        \includegraphics[width=.47\columnwidth]{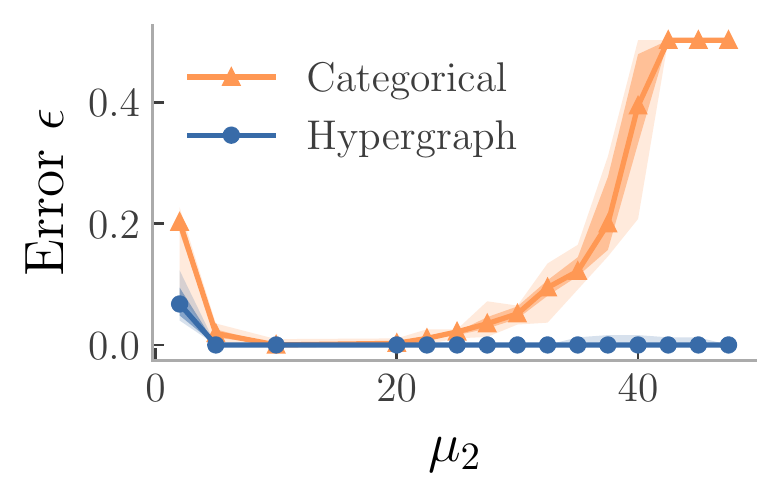}
    }\hfill
    \subfloat[\label{sfig:best_entropy_mu2}]{%
        \includegraphics[width=.47\columnwidth]{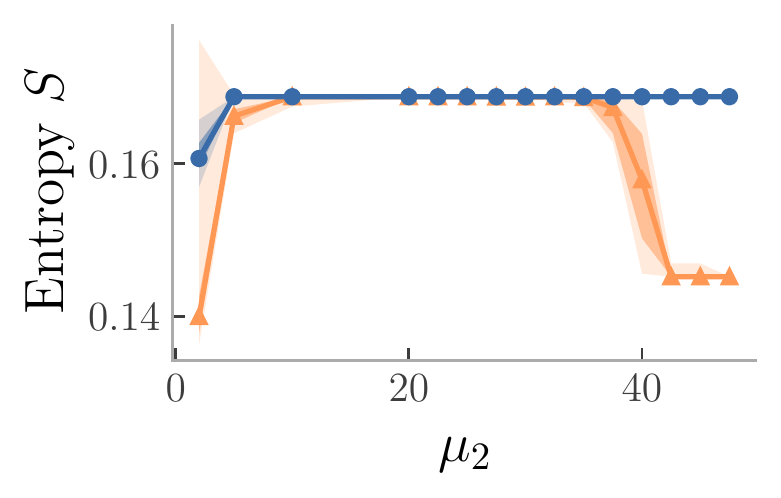}
    } \newline
    \subfloat[\label{sfig:best_residuals_mu2}]{%
    \includegraphics[width=\linewidth]{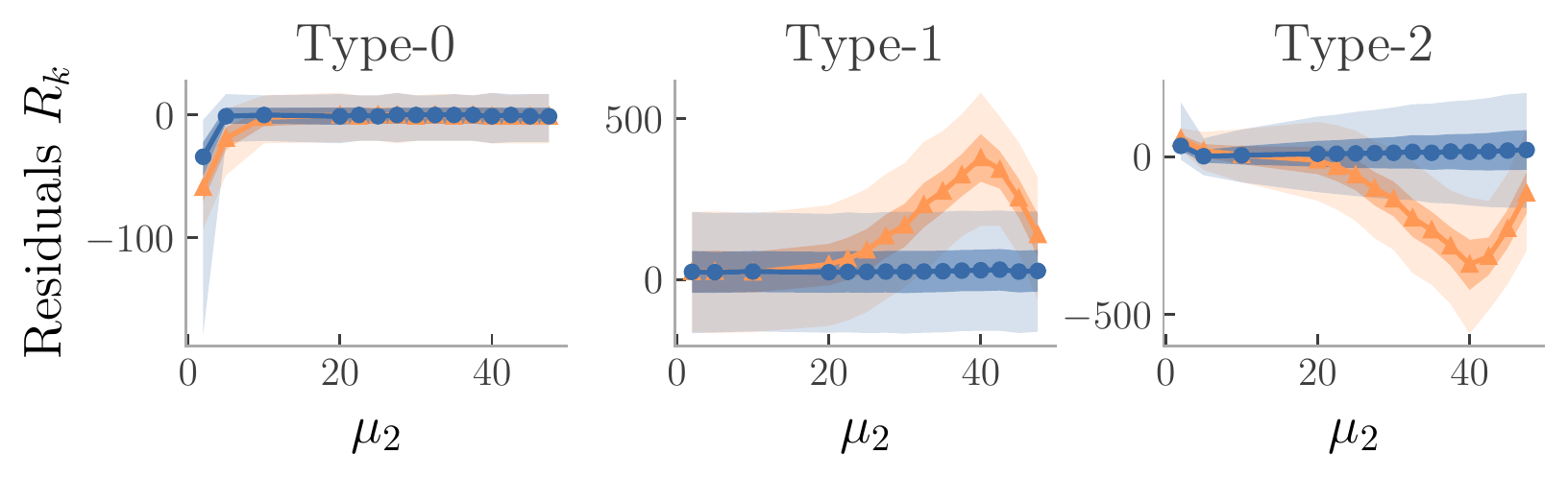}
    } \newline
    \subfloat[\label{sfig:best_confusion_matrix_mu2}]{%
    \includegraphics[width=\linewidth]{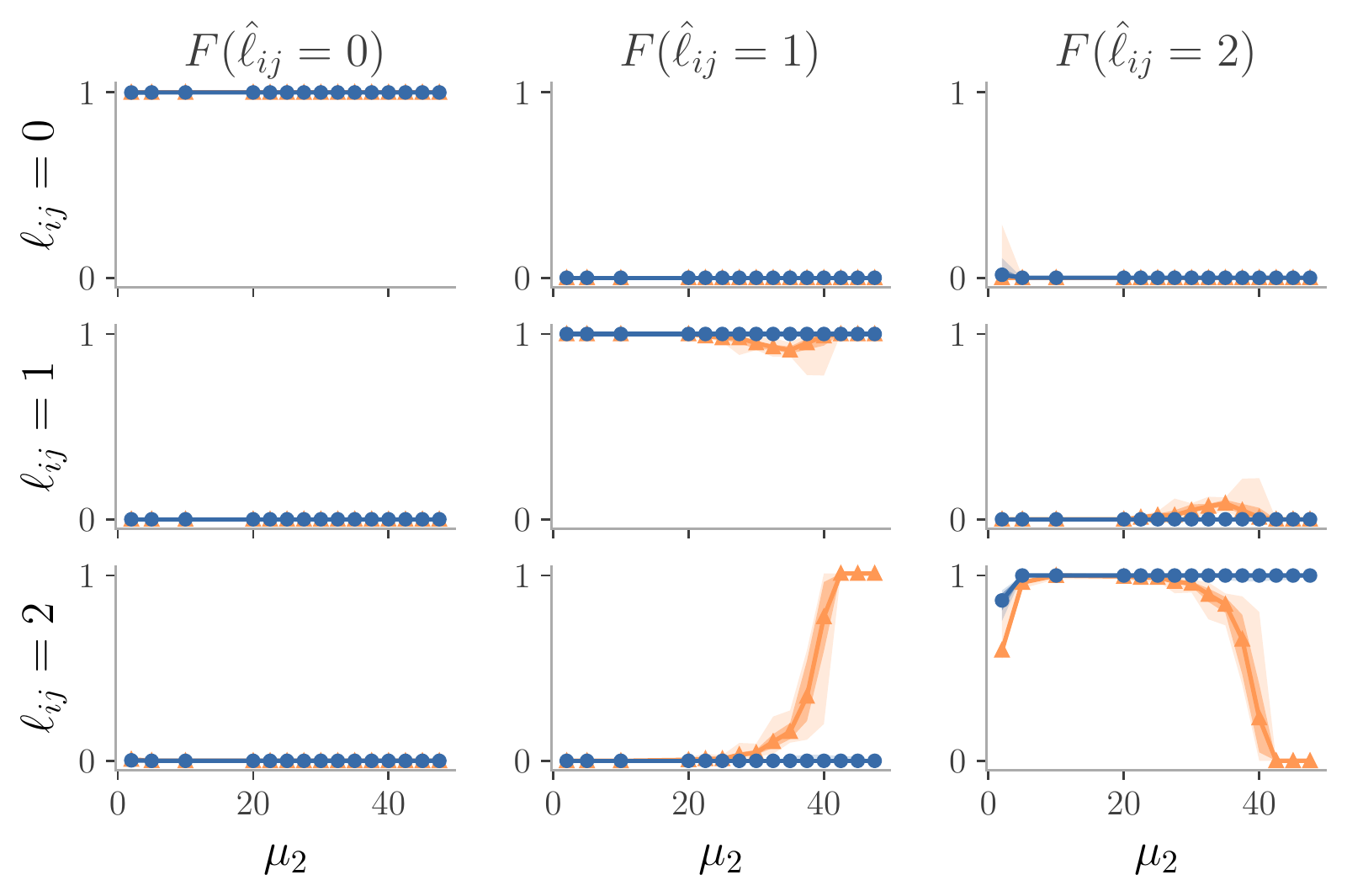}
    }
    \caption{
    Impact of the measurement rate ($\mu_2$) of type-2 interactions on the reconstruction of a best-case hypergraph.
    (a) Relative reconstruction error $\epsilon$. (b) Entropy $S$. (c) Sums of residuals $R_k$. (d) Normalized confusion matrix. The observations were generated with $\mu_0=0.01$, $\mu_1=50$ and various $\mu_2$ using the hypergraph model (blue) and the categorical edges graph model (orange). The hypergraph model displays (a, d) less misclassification errors (b) a larger entropy (c) lower residuals than the categorical edges graph model, which indicates a better reconstruction.
    See the caption of Fig.~\ref{fig:best_mu1} for details on the numerical experiment.
    }
    \label{fig:best_mu2}
\end{figure}

The MH algorithm is initialized at the ground truth hypergraph and parameters except in Table~\ref{tab:real_data_results} where it is initialized at a hypergraph with no 3-edge and 2-edges wherever $x_{ij} > 0$ and at parameters $\mu$ and $\phi_H$ set to the maximum likelihood estimator obtained from a Poisson mixture model. At each iteration, one of six possible moves is proposed:
\begin{enumerate}
  \item add $(a\!=\!1)$ a 2-edge with probability $\nu_2\eta$;
  \item remove $(a\!=\!0)$ a 2-edge with probability $\nu_2(1\!-\!\eta)$;
  \item add $(a\!=\!1)$ a 3-edge with probability $\nu_3\eta$;
  \item remove $(a\!=\!0)$ a 3-edge with probability $\nu_3(1\!-\!\eta)$;
  \item add $(a\!=\!1)$ a hidden 2-edge with probability $(1\!-\!\nu_2\!-\!\nu_3)\eta$;
  \item remove $(a\!=\!0)$ a hidden 2-edge with probability $(1\!-\!\nu_2\!-\!\nu_3)(1\!-\!\eta)$.
\end{enumerate}
We use $\eta=0.5$ and $\nu_2=\nu_3=0.4999$.

If the algorithm reaches a point where either no 2-edge or 3-edge can be added (removed), then one is removed (added) with probability 1.  If a move in which a hidden 2-edge should be added/removed has been chosen and that move is not possible (e.g. there are no hidden 2-edge to be removed), a completely new move is randomly chosen.

\begin{figure}[t]
    \subfloat[\label{sfig:worst_confusion_mu2}]{%
        \includegraphics[width=.47\columnwidth]{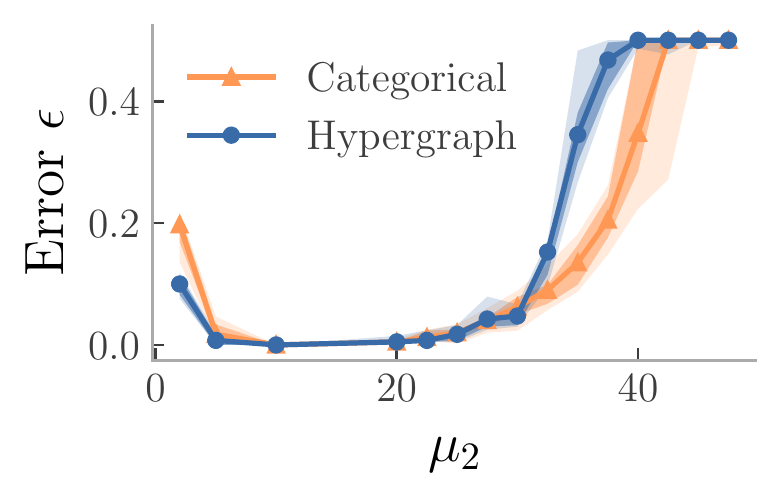}
    }\hfill
    \subfloat[\label{sfig:worst_entropy_mu2}]{%
        \includegraphics[width=.47\columnwidth]{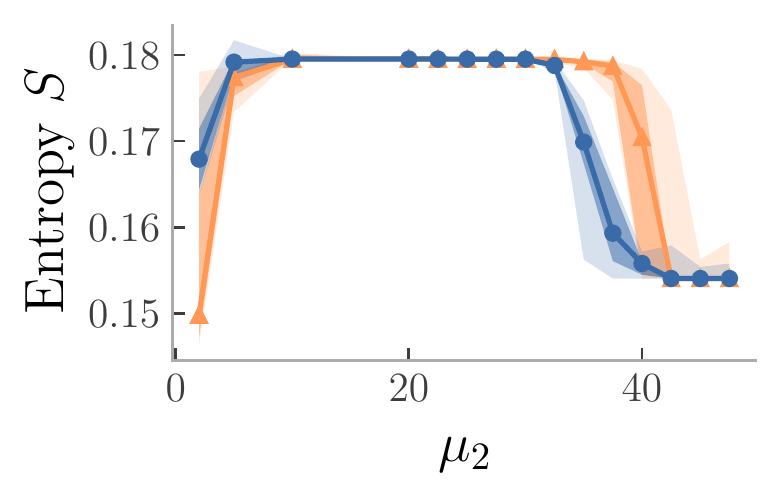}
    } \newline
    \subfloat[\label{sfig:worst_residuals_mu2}]{%
    \includegraphics[width=\linewidth]{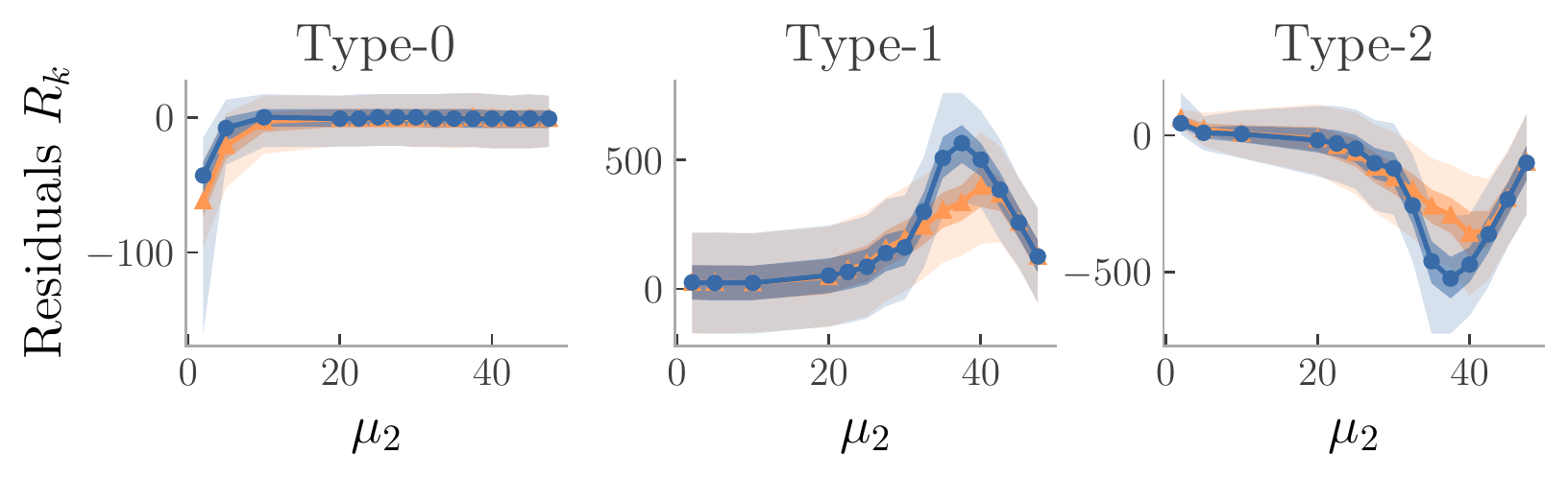}
    } \newline
    \subfloat[\label{sfig:worst_confusion_matrix_mu2}]{%
    \includegraphics[width=\linewidth]{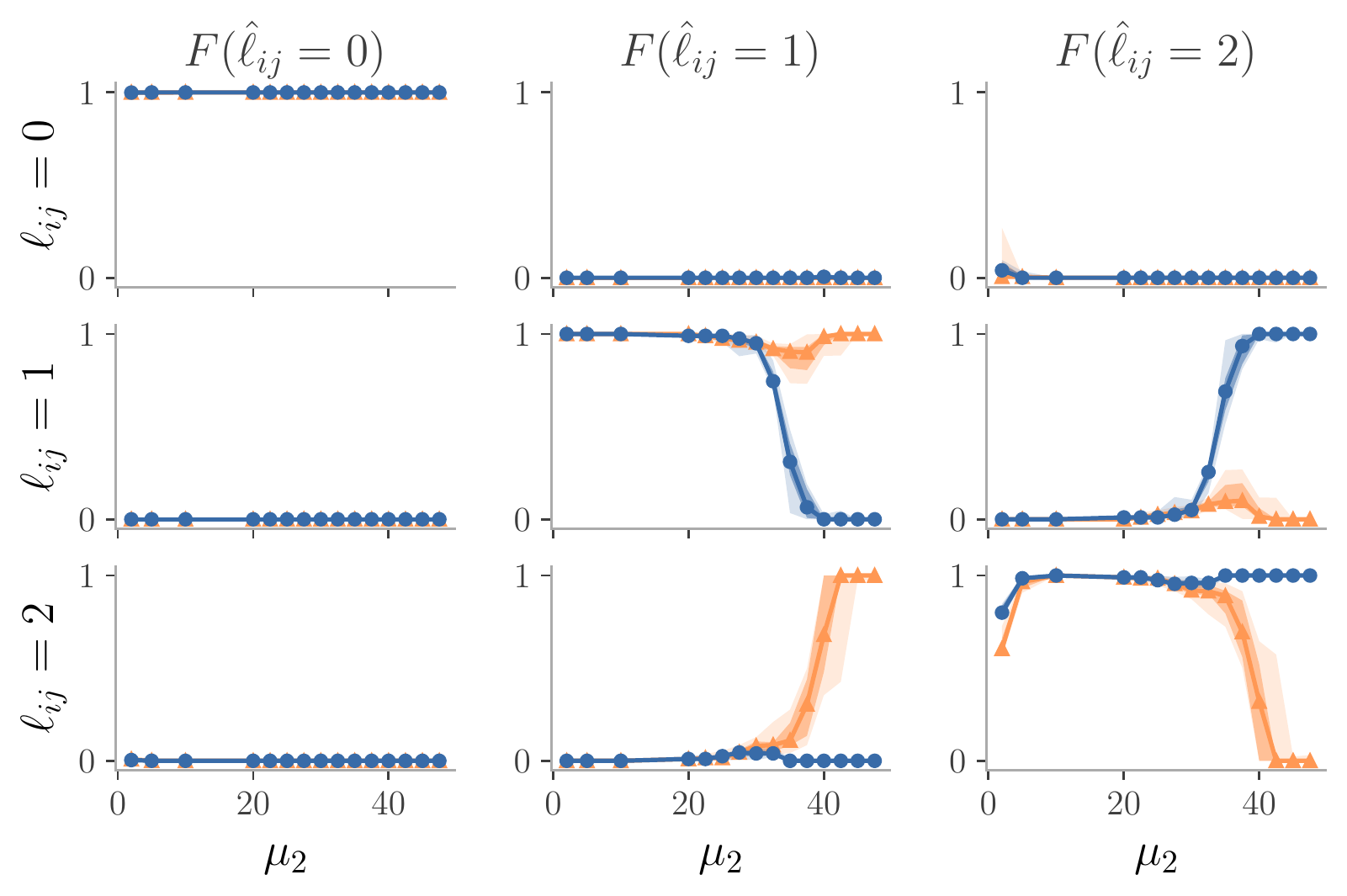}
    }
    \caption{
    Impact of the measurement rate ($\mu_2$) of type-2 interactions on the reconstruction of a worst-case hypergraph.
    (a) Relative reconstruction error $\epsilon$. (b) Entropy $S$. (c) Sums of residuals $R_k$. (d) Normalized confusion matrix. The observations are generated with $\mu_0=0.01$, $\mu_1=50$ and various $\mu_2$ using the hypergraph model (blue) and the categorical edges graph model (orange). The hypergraph model displays (a, d) more misclassification errors (b) a smaller entropy (c) greater residuals than the categorical edges graph model, which indicates a worse reconstruction.
    See the captions of Figs.~\ref{fig:best_mu1} and~\ref{fig:worst_mu1} for details on the numerical experiment.
    }
    \label{fig:worst_mu2}
\end{figure}

The proposed move, that would transform the hypergraph $H$ into a new one $H^*$, is accepted with probability
\begin{align}
    \alpha(H^*|H) = \min\qty(1,\ \frac{P(H^*, \theta|X)Q(H|H^*, X)}{P(H,\theta|X)Q(H^*|H, X)}).
    \label{eq:mh_hypergraph}
\end{align}
We now detail the proposal probability ratio $\frac{Q(H|H^*, X)}{Q(H^*|H, X)}$ for each of the 6 possible moves.

When a 3-edge is to be removed, it is chosen uniformly among the existing 3-edges.  When a 3-edge is to be added, the three vertices $(i,j,k)$ are chosen in three steps:
pick $i \sim P(i)$, pick $j \sim P(j|i)$ and pick $k \sim P(k|i)$ where
\begin{subequations}
\begin{align}
    P(i)   & = \frac{\sum_{l \neq i} (x_{il} + 1)}{\sum_r \sum_{s \neq r} (x_{rs} + 1)} \\
    P(j|i) & = \frac{x_{ij} + 1}{\sum_{l \neq i} (x_{il} + 1)}.
\end{align}
\end{subequations}
Since the order in which vertices are chosen does not matter, the probability that triplet $(i,j,k)$ is chosen is
\begin{multline}
    P(i,j,k) = 2 P(i) P(j|i) P(k|i)
             + 2 P(j) P(i|j) P(k|j) \\
             + 2 P(k) P(i|k) P(j|k).
\end{multline}
If this selection process results in the triplet $(i,j,j)$ or chooses an existing 3-edge, then the proposed move is automatically rejected since the distribution is only supported on \emph{simple} hypergraphs. Altogether, the proposal probability ratio for moves involving 3-edges can be summarized as
\begin{align}
    \frac{Q(H|H^*, X)}{Q(H^*|H, X)} = \qty(\frac{1}{\eta P(i,j,k)} \frac{1-\eta}{h_2 + a})^{2a-1}.
\end{align}

\begin{figure}[t]
    \centering
    \includegraphics[width=\linewidth]{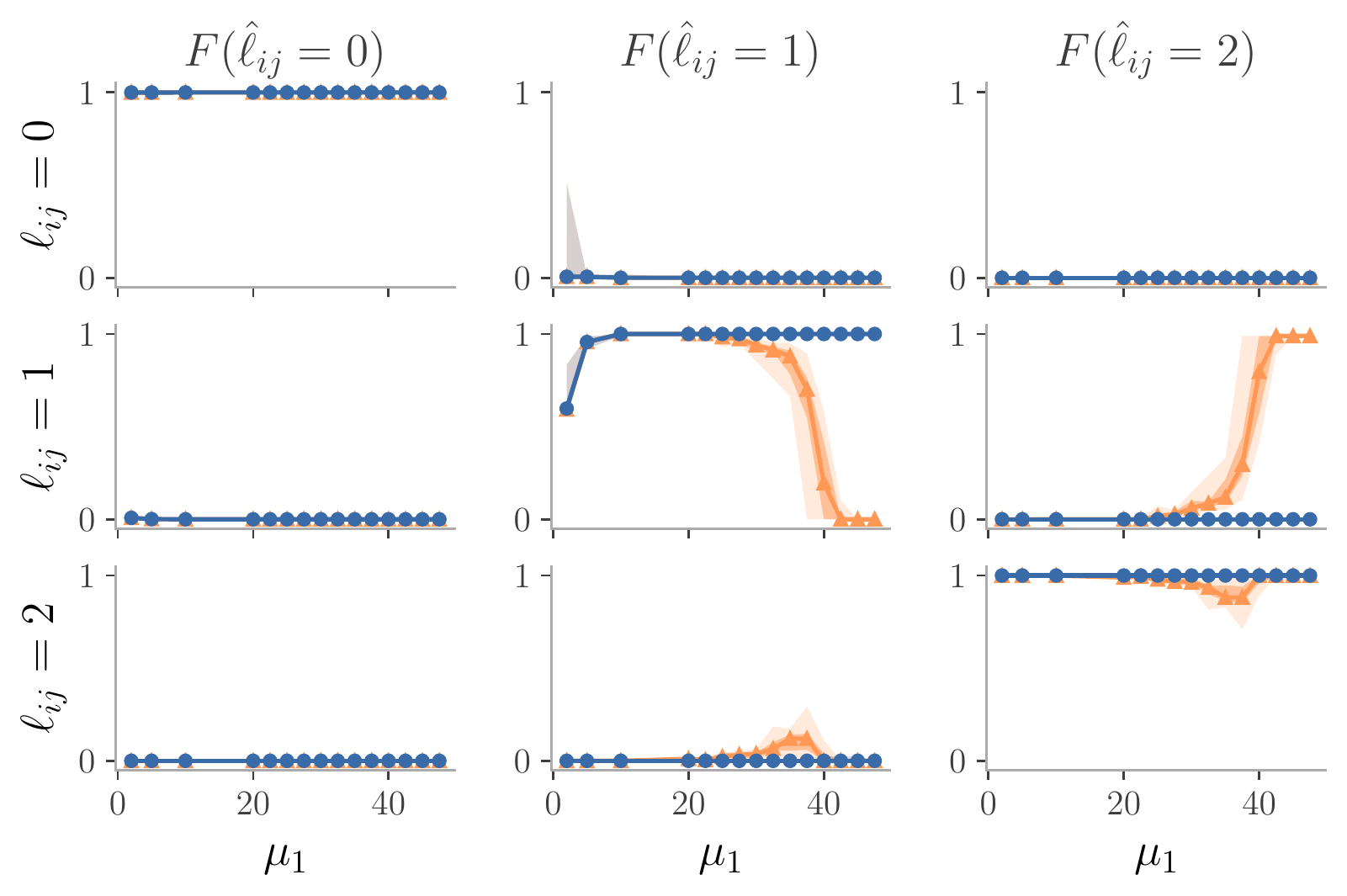}
    \caption{
    Normalized confusion matrix associated to the simulation of Fig.~\ref{fig:best_mu1}. The categorical edges graph model favors the strong edges when $\mu_1$ approaches $\mu_2$, which leads to an inferior reconstruction compared to the hypergraph model that commits little to no error.
    }
    \label{fig:best_confusion_matrix_mu1}
\end{figure}

When a 2-edge needs to be removed, it is chosen uniformly among the existing 2-edges. When a 2-edge $(i,j)$ needs to be added, it is chosen proportionally to the weight
\begin{align}
    \omega_{ij} = \begin{cases}
        x_{ij}+1 & \text{if }(i,j) \not\in E \\
        0 & \hfil\text{otherwise}.
    \end{cases}
\end{align}
Altogether, the proposal probability ratio for moves involving 2-edges can be summarized as
\begin{multline}
    \frac{Q(H|H^*, X)}{Q(H^*|H, X)}
    = \qty( \frac{\sum_{r<s} \omega_{rs} + a(x_{ij}+1)}{\eta (x_{ij}+1)} \frac{1-\eta}{h_1+a} )^{2a-1}
\end{multline}

Our definition of the types of interactions $\ell_{ij}$ [Eq.~\eqref{eq:phg_labels}] implies that hidden 2-edges do not contribute to the likelihood [Eq.~\eqref{eq:likelihood}]; their addition/removal depends solely on the hypergraph model.
However, the \emph{cost} of removing a 3-edge depends on the number of hidden 2-edges underneath. Because of this asymmetry, we found that running the MH algorithm with the four previous moves tend to get stuck with certain configurations of hidden 2-edges. Our solution has been to propose two additional moves specifically targeting hidden 2-edges.

To propose the addition/removal of a hidden 2-edges, we first regroup every existing hidden 2-edges into a set $C_0$ and every ``non-existing'' hidden 2-edges into a set $C_1$. (These non-existing hidden 2-edges are interactions of type 2 for which the corresponding 2-edge does not exist.) We then draw the number $m$ of 2-edges to add/remove from a truncated geometric distribution of parameter $\chi_a$ on the interval $[2, |C_a|]$. If $|C_a| < 2$, the move is automatically rejected. We force $m \geq 2$ to ensure these two additional moves do not overlap with the previous two moves involving 2-edges in the MH algorithm acceptance probabilities. Finally, we choose uniformly $m$ hidden 2-edges in $C_a$ and store them in the set $e$; their addition/removal consist in the proposed move. The probability of a given set $e$ is
\begin{align}
    P(e|C_a, \chi_a) = \frac{(1-\chi_a)^{|e|-2}\chi_a}{1-(1-\chi_a)^{|C_a|-1}} \binom{|C_a|}{|e|}^{-1},
\end{align}
and the proposal probability ratio for moves involving hidden 2-edges only is
\begin{align}
    \frac{Q(H|H^*, X)}{Q(H^*|H, X)}
    = \left( \frac{1\!-\!\eta}{\eta} \right)^{2a-1} \frac{P(e|C_{1-a} \bigcup e, \chi_{1-a})}{P(e|C_a, \chi_a)}.
\end{align}
In the simulations, we use $\chi_0=0.99$ and $\chi_1=0.01$ as we want to remove more frequently than add hidden 2-edges.

\begin{figure}[t]
    \centering
    \includegraphics[width=\linewidth]{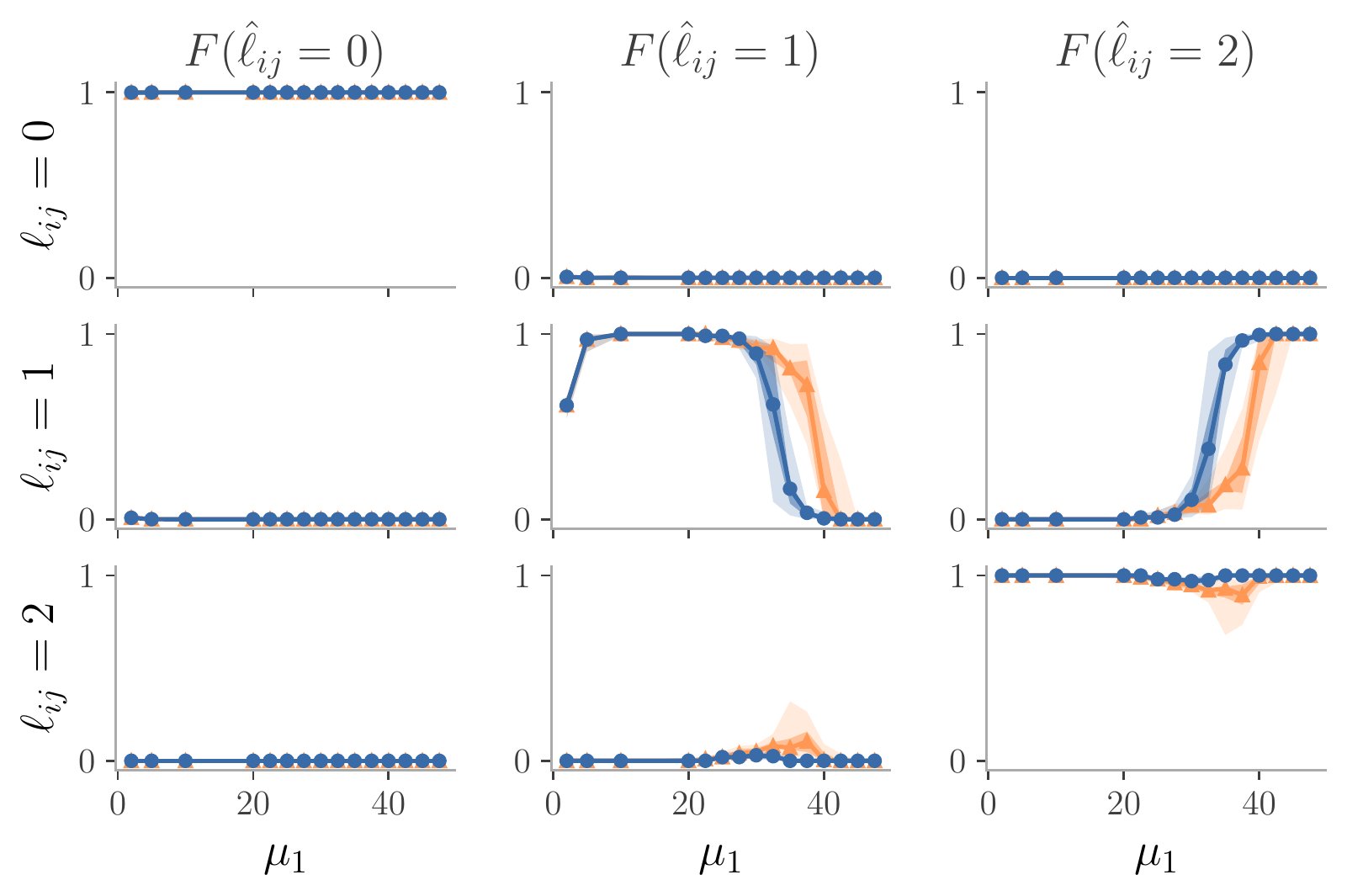}
    \caption{
    Normalized confusion matrix associated to the simulation of Fig.~\ref{fig:worst_mu1}. While the categorical edges model still favors strong edges to weak edges, the hypergraph model favors more strongly 3-edges and displays a worse performance for the worst-case hypergraph.
    }
    \label{fig:worst_confusion_matrix_mu1}
\end{figure}

\subsection{Convergence}

We stop the two previous MH algorithms whenever the likelihood stabilizes, meaning the chains have reached stationarity. We consider that this has happened when the relative change in the average likelihood of the last $W$ iterations is smaller than a tolerance parameter $\delta$. We use $W=20000$ and $\delta=0.02$ in our simulations.

Further, to ensure the MH algorithm runs long enough but not too long, we set a minimum $I_\text{min}$ and maximum $I_\text{max}$ number of iterations. We adjust these values  empirically with a test run, but they are roughly $I_\text{min}=200000$ and $I_\text{max}=1000000$. Finally, for each posterior distribution sample, we run four chains and keep the one with the highest average likelihood.

\section{Regime \texorpdfstring{$\mu_1>\mu_2$}{mu1>mu2} and confusion matrices}

We mentioned in Sec.~\ref{ssec:s_models} that conditions such as  $\mu_1<\mu_2$ or $\mu_1>\mu_2$ need not be imposed in the prior distributions since 2-edges and 3-edges are fundamentally different. As a complement to the analysis presented in Sec.~\ref{ssec:synthetic_inference}, we investigate the case where $\mu_2$ is varied between $\mu_0=0.05$ and $\mu_1=50$.

Comparison between Figs.~\ref{fig:best_mu1} and~\ref{fig:worst_mu1} and Figs.~\ref{fig:best_mu2} and~\ref{fig:worst_mu2} suggests that both scenarios are quite similar, as expected. In particular, note that the apparent swap in the sums of residuals for the categorical edges graph model is simply due to the redefinition of $\ell_{ij}$ to accommodate the restriction that $\mu_1<\mu_2$ in the model. Indeed we redefine
\begin{align}
    \ell_{ij} = \begin{cases}
        1 & (i,j) \in E_2,\\
        2 & (i,j) \in E_1,\\
        0 & \hfil\text{otherwise.}
    \end{cases}
\end{align}

The only noteworthy difference between the two sets of simulations occurs when $\mu_2$ approaches $\mu_0$. We observe the same phenomenon than when $\mu_1$ approaches $\mu_2$ from the left: the information in the neighborhood used by the hypergraph model allows for a more accurate reconstruction (i.e., smaller $\epsilon$, larger $S$).
Interestingly, this effect is also apparent in the worst-case hypergraphs.
Because the structures we consider are sparse, many non-interacting pairs form triangles such that when $\mu_2$ approaches $\mu_0$, they can easily be confused with 3-edges.

Figures~\ref{fig:best_confusion_matrix_mu1}, \ref{fig:worst_confusion_matrix_mu1}, \ref{sfig:best_confusion_matrix_mu2} and~\ref{sfig:worst_confusion_matrix_mu2} show the normalized confusion matrix for the best-case and worst-case hypergraphs. The entries of the normalized confusion matrix $\tilde{c}_{rs}$ are the proportion of interactions of type $\ell_{ij}=r$ that were predicted as $\hat \ell_{ij}=s$ by the model
\begin{align}
    \tilde{c}_{rs} = \frac{c_{rs}}{c_{r0} + c_{r1} + c_{r2}}.
\end{align}
For instance, the element $\tilde{c}_{21}$ is the proportion of projected 3-edges predicted as 2-edges in the hypergraph model.

When the categorical edges graph model ends up inferring only one type of interaction, there are two equivalent reconstructed graphs: all interactions are weak edges or all interactions are strong edges. Noting that in less extreme cases, the model naturally favors strong edges due to the larger associated variance in the likelihood, we set all interactions to strong edges whenever classifies them all as a weak edges.

We see that for the best-case structure in Figs.~\ref{fig:best_confusion_matrix_mu1} and~\ref{sfig:best_confusion_matrix_mu2}, the hypergraph model makes little to no error. As we increase $\mu_1$, we also observe a gradual increase of the number of misclassified weak edges for the categorical edges model. For the worst-case structure, the results in Figs.~\ref{fig:worst_confusion_matrix_mu1} and~\ref{sfig:worst_confusion_matrix_mu2} show the the hypergraph model favors 3-edges and that the categorical edges model favors strong edges.

\end{document}